\documentstyle[preprint,prc,aps,floats,epsf]{revtex}

\tightenlines

\begin{document}

\def\tauvec{{\bbox{\tau}}}
\def\grad{{\bbox{\nabla}}}
\def\wt{\widetilde}
\def\comment#1{[[{\it #1?}]]}        
\def\scalar{\rho_{\scriptscriptstyle S}}
\def\scalarN{{\wt\scalar}}
\def\vector{j_{\scriptscriptstyle V}}
\def\vectorN{\wt \vector}
\def\isovectorVector{j_{\tauvec}}
\def\isovectorVectorN{\wt j_{\tauvec}}
\def\tensor{s}
\def\tensorN{\wt\tensor}
\def\isovectorTensor{s_{\tauvec}}
\def\isovectorTensorN{\wt\isovectorTensor}
\def\del{\partial}
\def\delN{{\wt\partial}}
\def\kfermi{{\null k_{\scriptscriptstyle F}}}
\def\opt#1{\wt\Omega_{#1}}
\def\optder#1{\wt\Delta_{#1}}
\def\optalpha#1{\wt A_{#1}}
\def\project{{\cal P}}
\def\fpi{f_{\pi}}
\def\Mstar{M^\ast}

\draft

\preprint{OSU--97-220}

\title{Relativistic Point-Coupling Models as 
\\ Effective Theories of Nuclei}

\author{John J. Rusnak and R. J. Furnstahl}
\address{Department of Physics \\
         The Ohio State University,\ \ Columbus, Ohio\ \ 43210}
%

%
\date{August, 1997}

\maketitle
\begin{abstract}
Recent studies have shown that concepts of effective field
theory such as naturalness can be 
profitably applied to relativistic mean-field models of nuclei.
Here the analysis by Friar, Madland, and Lynn of naturalness in
a relativistic point-coupling model is extended.
Fits to experimental nuclear data support naive dimensional analysis
as a useful principle
and imply a mean-field expansion analogous to that found
for mean-field meson models.
\end{abstract}

\section{Introduction}

Quantum Chromodynamics\ (QCD) is believed to be the underlying theory
of hadrons and their interactions
and therefore of nuclei as well.
However, {\em direct}\/ solutions of QCD
for nuclei (e.g., lattice calculations) are not presently feasible.
On the other hand, the modern perspective of effective field theory (EFT)
\cite{Weinberg1,Georgi2,LePage,EckerA,WEINBERG95,Kaplan}
provides a framework  for constructing
theories based on observed hadronic degrees of freedom
that can faithfully reproduce low-energy QCD.
An example of a successful EFT  is chiral perturbation 
theory (ChPT)\cite{Weinberg1,Gasser1,Meisner1},
which systematically describes low-energy hadronic processes in the vacuum.

The success of effective chiral lagrangians in the vacuum 
has motivated the application of EFT concepts to models 
of nuclear properties.
A chiral effective lagrangian for nuclei
was presented in Ref.~\cite{Tang}, and
has lead to new insights into the successes of
quantum hadrodynamic (QHD) models\cite{Serot,SEROTb} of nuclei. 
In that work, a general model in which  nucleons interact via 
mesonic degrees of freedom was constructed and applied to nuclei at 
the one-baryon-loop level (Hartree approximation).

The effective lagrangian and energy functional were organized 
according to Georgi's naive dimensional analysis 
(NDA) \cite{GeorgiAndMan,Georgi}, 
which
provides a controlled mean-field expansion in terms of powers and
derivatives of the mean fields {\it if\/} the dimensionless
coefficients identified in the NDA are of order one.
The latter property is called ``naturalness'' in this context.
Detailed fits to nuclear observables validated this expansion and
truncation scheme \cite{Tang}.
This result is somewhat surprising since the NDA counting
is based on absorbing short-distance physics into the coefficients
of the effective lagrangian, but a mean-field functional
fit to nuclear data must also absorb long-distance many-body
effects.

A useful interpretation of the finite-density mean-field effective
theory is in terms of density functional theory 
(DFT) \cite{DREIZLER90,SPEICHER92,SCHMID95,ENGEL95}.
In a DFT formulation of the relativistic many-body problem, 
one works with an energy functional of scalar densities and vector
four-currents.
Minimization of the functional gives rise to variational equations
that determine the ground-state densities.
By introducing a complete set of Dirac wave functions, one can
recast these variational equations as Dirac equations for occupied
orbitals; the single-particle hamiltonian contains
{\it local\/} scalar and vector potentials, not only in the Hartree
approximation, but in the general case as well.
The Hartree approximation only limits the form
of the potentials.
Thus the effects of many-body physics such as short-range correlations
are incorporated (approximately) by a direct fit to nuclear observables.

In mean-field meson models, the scalar and vector meson fields play
the role of auxiliary Kohn-Sham potentials.
Alternatively, one can expand the local potentials directly in terms
of nucleon densities.
The corresponding lagrangian replaces meson exchange with point-coupling
(contact) interactions.
In Ref.~\cite{Madland}, such a relativistic 
point-coupling lagrangian was introduced and applied to nuclei at the 
mean-field level.
Subsequently,
Friar et al.~\cite{Friar} examined this model
for naturalness and concluded that the parameters were, in fact, 
mostly natural.
Here, the analysis of point-coupling models in the context of
EFT's is reexamined and
broadened in a manner consistent with the analysis of
the chiral mean-field meson model of Ref.~\cite{Tang}.

We find that the assumption of naturalness is justified and
that the density expansion implied
by NDA is applicable.
In principle, the parameters found here
could be related to those in the meson 
model of Ref.~\cite{Tang} by using the equations of motion in the
lagrangian of \cite{Tang} to systematically eliminate the meson fields.
However, the point-coupling parameters are underdetermined by the fits,
which limits the usefulness of the comparisons.
New optimization procedures described below
suggest that a different organization that
exploits cancellations characteristic of relativistic models may
be more productive.

\section{The Point-Coupling Lagrangian}

At present we cannot derive an effective 
point-coupling lagrangian directly
from QCD.  In the same spirit as ChPT and the effective meson
model of Ref.~\cite{Tang}, a general point-coupling effective lagrangian  
is therefore constructed consistent with
the underlying symmetries of QCD (e.g., Lorentz covariance,
gauge invariance, and  chiral symmetry).
In this work, we construct a one-loop energy-functional from the
lagrangian and determine the couplings
by fits to nuclear observables.
As noted above and discussed in Ref.~\cite{SEROTb},
this approach approximates a density functional that, if
sufficiently general, incorporates many-body effects beyond the Hartree 
approximation.
We expect this approximation to be reasonable because of the large
scalar and vector potentials (``Hartree dominance'') \cite{Tang,SEROTb}. 

This framework will only be useful if we can identify a valid
expansion and truncation scheme.
This requires an organization of terms in the effective lagrangian
and a way to estimate the couplings.
While precise relations between these  couplings 
 and the underlying QCD parameters are unknown,
an estimate of the magnitude of the couplings
can be obtained by applying 
Georgi's naive dimensional analysis (NDA)\cite{GeorgiAndMan,Tang,FRIAR96b}.

The procedure is to extract from each term in the lagrangian 
the dependence on two primary 
physical scales of the effective theory, 
the pion decay constant, $f_\pi \approx 94\,$MeV
and
a larger mass scale, $\Lambda\sim 4\pi f_\pi/\sqrt{N_f}$ (where $N_f$ is 
the number  of light flavors) \cite{GeorgiAndMan,Georgi}.
The scale $\Lambda$ is associated with the 
mass scale of physics beyond the Goldstone bosons (pions):
the non-Goldstone boson masses or the nucleon mass.
This mass scale ranges from the scalar mass\ $\approx 500\,$MeV to
the baryon mass $\approx 1\,$GeV.  
When a specific value is needed in this work,
$\Lambda$ will be taken to equal
the $\rho$-meson mass ($770\,$MeV), roughly in the center of this range.
To establish the canonical normalization of the strongly interacting
fields, 
an inverse factor of $f_{\pi}$ is included 
for each field  and an
overall factor of $f_{\pi}^{2}\Lambda^{2}$ fixes the normalization
of the lagrangian.
The physics of NDA is discussed further in Refs.~\cite{Tang} and \cite{SEROTb}.

We construct the point-coupling effective lagrangian
as an expansion in powers of the nucleon
scalar, vector, isovector-vector, tensor, and isovector-tensor densities
scaled according to NDA:
\begin{eqnarray}
\scalarN &\equiv& {\scalar\over f_\pi^2\Lambda}\equiv
{\overline N N\over f_\pi^2\Lambda}\ ,  \label{eq:eqone}\\
\vectorN^\mu &\equiv & {\vector^\mu\over f_\pi^2\Lambda}
	\equiv{\overline N \gamma^\mu N\over f_\pi^2\Lambda}\ ,\\
\isovectorVectorN^\mu &\equiv &{\isovectorVector^\mu\over f_\pi^2\Lambda} 
	\equiv{\overline  N\gamma^\mu{\tauvec\over 2}N
		\over f_\pi^2\Lambda}\ ,\\
\tensorN^{\mu\nu} &\equiv & {\tensor^{\mu\nu}\over f_\pi^2\Lambda} \equiv
	{\overline N\sigma^{\mu\nu} N\over f_\pi^2\Lambda}\ ,\\
\isovectorTensorN^{\mu\nu} &\equiv & 
	{\isovectorTensor^{\mu\nu}\over f_\pi^2\Lambda}
	\equiv {\overline N \sigma^{\mu\nu}{\tauvec\over 2}N \over
		f_\pi^2\Lambda}\ ,  \label{eq:eqfive}
\end{eqnarray}
where $N$ is the nucleon field.
The lagrangian is also organized according to an expansion in derivatives
acting on these densities; NDA dictates that 
each derivative is scaled by
$\Lambda$:
\begin{eqnarray}
\delN^\mu&\equiv&{\del^\mu\over \Lambda}\ . \label{eq:eqsix}
\end{eqnarray}
Except for the kinetic term,
derivatives acting on an individual nucleon field,  
rather than on a
density, are eliminated by field redefinitions as in Ref.~\cite{Tang}
in favor of terms of the form of Eqs.~(\ref{eq:eqone})--(\ref{eq:eqfive}).
This is because time derivatives on the nearly on-shell valence
nucleons are of $O(M)$ and are therefore not suppressed according to
(\ref{eq:eqsix}).
As discussed in Ref.~\cite{Tang}, this procedure is only
able to eliminate derivatives in the combination
$\gamma^\mu \partial_\mu$ (and therefore $\partial^\mu\partial_\mu$).
We can transform away mixed-derivative terms such as
$(\overline N\partial_\mu N)(\overline N\partial^\mu N)$
in favor of tensor terms, which are then neglected (see below) \cite{Tang}.

Naive dimensional analysis 
provides an organizational principle that directly translates
into numerical estimates at the mean-field level.
For example, each additional power of $\rho_s$ is accompanied by
a factor of $\fpi^2\Lambda$.
The ratios of scalar and vector densities to this factor at nuclear matter
equilibrium density are between 1/4 and 1/7 \cite{FRIAR96b}, which 
serves as an expansion parameter. 
Similarly, one can anticipate good convergence
for gradients of the densities, since the relevant scale for
derivatives in finite nuclei should be roughly the nuclear surface
thickness $\sigma$, and so the dimensionless expansion parameter
is $1/\Lambda\sigma \le 1/5$.
This expansion is only useful, however, if the coefficients are
not too large.

In effective lagrangians of QCD applied to scattering problems, 
fits to experimental data suggest that when
NDA is applied the remaining  
dimensionless coefficients are of order unity. 
This is known as ``naturalness''; it is an essential feature if
NDA is to be useful as an organizational scheme.  
The premise of naturalness in point-coupling models
is to be tested here for {\em finite-density} applications through
fits to experimental nuclear data.
Naive dimensional analysis
applied to the point-coupling model of Nikolaus et~al.~\cite{Madland,Friar}
and to the meson model of Ref.~\cite{Tang} already suggest
that effective nuclear models are natural,
which in turn implies
a convergent mean-field expansion in density based on an organization
prescribed by NDA \cite{Tang}.
This is a nontrivial result, because the naturalness assumption implies
that all the short-distance  physics  (with scale $1/\Lambda$) is incorporated
into the coefficients of the effective lagrangian, while long-distance
finite-density effects should be calculated explicitly in a systematic
application of the effective lagrangian.
At the mean-field level, however, we also approximately absorb long-distance
many-body effects from ladder and ring diagrams \cite{SEROTb}.

The effective lagrangian should in principle
contain every possible term (allowed by symmetries)
to a given order under this organization.
However, certain classes of terms will be poorly determined by fits
to bulk nuclear observables.
Here meson-exchange phenomenology and experience with relativistic
mean-field meson models are useful guides.
Thus, we follow the physics motivation of Ref.~\cite{Tang}
to determine which terms will be essential and which can be omitted.
In particular,
each term we include corresponds to one in the meson-nucleon 
lagrangian, as identified 
through a simple leading-order analysis.
For example, at leading order, the scalar field, $\phi$, in the meson model 
of Ref.~\cite{Tang} is proportional to
the scalar density, $\scalarN$;
terms second order in the scalar density here (including those
containing derivatives) are therefore related to the mass and kinetic terms
of the scalar meson field 
as well as its Yukawa coupling to the nucleon in the meson model.
The term cubic in the scalar density
in the point-coupling model 
has a correspondence with the term cubic in the scalar field, and so on.  
The {\em absence} of a tensor boson in the mesonic model 
corresponds to the absence of terms of the form  
$[\overline N\sigma^{\mu\nu} N]^2$ and
$[\delN_\alpha\overline N\sigma^{\mu\nu} N]^2$ 
in our point-coupling lagrangian.
Tensor meson masses are large and tensor mean-field densities are small,
so even if the coefficients are natural, we expect that
such terms would have a small effect. 
An 
analysis of terms not included here is postponed
to a future investigation.

The resulting point-coupling
lagrangian is divided into four parts:
\begin{eqnarray}
{\cal L} & = & {\cal L}_{\rm NN} + {\cal L}_{\pi} + {\cal L}_{\rm EM}
+ {\cal L}_{\rm VMD} \ . 
\end{eqnarray}
The pure nucleon contact interactions are contained in ${\cal L}_{\rm NN}$,
and take the form
\begin{eqnarray}
{\cal L}_{\rm NN} &=& \overline N[i\partial^\mu\gamma_\mu-M] N
\nonumber \\
&&-f_\pi^2\Lambda^2\biggl\{ \ 
\scalarN^2 \Bigl[
      \wt\kappa_2 + \wt\kappa_3\scalarN + \wt\kappa_4\scalarN^2 \Bigr]
     \nonumber\\
 &&\qquad\qquad
    \null + (\vectorN^\mu)^2 \Bigl[\wt\zeta_2 + \wt\eta_1\scalarN
        +\wt\eta_2\scalarN^2
         +\wt\zeta_4(\vectorN^\nu)^2 \Bigr]
       \nonumber\\
 &&\qquad\qquad
  \null + [\delN^\mu\scalarN]^2 \Bigl[\wt\kappa_d 
      + \wt\alpha_1 \scalarN \Bigr]
  + [\delN^\mu\vectorN^\nu]^2 \Bigl[\wt\zeta_d + \wt\alpha_2 \scalarN \Bigr]
\nonumber\\
 &&\qquad\qquad
 \null + [\isovectorVectorN^\mu]^2 \Bigl[\wt\xi_2 + \wt\eta_\rho \scalarN \Bigr]
   +  {\wt\xi}_d [\delN^\mu\isovectorVectorN^\nu]^2
        \nonumber\\
  &&\qquad\qquad
 \null + \wt f_\rho
	\delN^\mu\isovectorVectorN^\nu\cdot{\isovectorTensorN}{}_{\mu\nu}
    +  \wt f_v\delN^\mu\vectorN^\nu
        \tensorN_{\mu\nu}\biggr\}\ .  \label{eq:LNN}
\end{eqnarray}
The terms are organized to manifest the expansion and truncation of
${\cal L}$ in powers and derivatives of the densities. 
The 
point-coupling parameters have a ``tilde'' over them to distinguish
them from the corresponding parameters of the  hadron lagrangian 
in Ref.~\cite{Tang}.
We follow Ref.~\cite{Madland} by not including counting factors in
applying NDA, although they were included in Ref.~\cite{Tang}.
This prescription will be tested empirically by our fits.

Compared to the point-coupling lagrangian of Nikolaus et al.~\cite{Madland},
our lagrangian includes additional terms with coefficients
$\wt\eta_{1}$, $\wt\eta_{2}$, $\wt\alpha_{1}$,
$\wt\alpha_{2}$, $\wt f_{\rho}$, and
$\wt f_{v}$.
We also exclude contributions that would correspond to 
an isovector, scalar channel in the meson lagrangian.  
This meson was not included in Ref.~\cite{Tang} 
based on meson-exchange phenomenology,%
\footnote{The NN interaction in that channel is weak \cite{Machleidt}.
  There is no meson with these quantum numbers with a mass
  below 1~GeV, and two (identical) pions in a $J=0$ state cannot
  have $T=1$, so there is no analog to the $\sigma$ meson.}
 and we note that 
the corresponding point-coupling coefficient in \cite{Madland} was found to be
unnaturally small.
Finally, we also postpone consideration of terms with an 
explicit dependence on the four-velocity $u^\mu$ of the nuclear medium
(such as $u_{\mu}\vectorN^{\mu}$), which might arise in an effective mean-field
energy functional.

The electromagnetic kinetic and interaction terms are contained
in ${\cal L}_{\rm EM}$ and ${\cal L}_{{\rm VMD}}$. 
Electromagnetic observables are calculated as an expansion 
in the electric charge $e$, 
as well as a derivative expansion.
Here we work
to first order in $e$; thus, only couplings to the nucleon that
are linear in the photon field are considered.
The lowest order terms in a derivative expansion are contained
in ${\cal L}_{\rm EM}$ and take the same form as the photon-nucleon
coupling terms in the meson model of Ref.~\cite{Tang}:
\begin{eqnarray}
{\cal L}_{\rm EM}&=&-{e\over 2}\overline N A^\mu\gamma_\mu(1+\tau_3) N
-{e\over 4M}F_{\mu\nu}\overline N\lambda\sigma^{\mu\nu} N\nonumber\\
&&-{e\over 2M^2}\partial_\nu F^{\mu\nu}\overline N(
	\{\wt\beta_s+\wt\beta_v\tau_3\}\gamma_\mu) N
-{1\over 4}F^{\mu\nu} F_{\mu\nu}\ .
\end{eqnarray}
The anomalous magnetic moment, $\lambda$, is given by 
\begin{eqnarray}
\lambda = {1\over 2}\lambda_p(1+\tau_3) + {1\over 2}\lambda_n(1-\tau_3)\ ,
\end{eqnarray} 
where $\lambda_{p}$ and $\lambda_{n}$ are the anomalous moments for
the proton and neutron.

In a mesonic model,
an adequate description of the low-momentum electromagnetic 
form factors of a single nucleon 
is achieved through a combination of
vector-meson dominance (VMD) \cite{VMD,BROWN}  and direct
couplings of the photon to nucleons \cite{Tang,BROWN}. 
Since heavy-meson degrees of freedom are eliminated here in favor of
contact interactions, these effects 
must be expressed in the point-coupling model
through direct higher-derivative couplings of the photon
to the nucleon, contained in ${\cal L}_{\rm VMD}$.
In practice we only include the next correction to the
charge form factors:
\begin{eqnarray}
{\cal L}_{\rm VMD}&=&-{e\over M^4}
\partial^2\partial_\nu F^{\mu\nu}\overline N(\{\wt\delta_{s}
	+\wt\delta_{v}\tau_3\}\gamma_\mu) N\  .\label{eq:VMD}
\end{eqnarray}
The coefficients of ${\cal L}_{\rm EM}$ and ${\cal L}_{\rm VMD}$
are fixed by experimental electromagnetic  data for protons
and neutrons in the vacuum 
and are not varied independently when fitting to nuclear
data.  This is detailed below.
The end result is that the composite electromagnetic structure of the nucleons
is incorporated as a derivative expansion, which is appropriate for
low-momentum physics.
In the point-coupling model the expansion is constructed explicitly
(at tree level) to a given order in momentum.  In a meson model, vector
meson dominance incorporates contributions to the form factor
to all orders in the momentum.

The pion kinetic and interaction terms are contained in ${\cal L}_\pi$.
The procedure for constructing this part of the lagrangian
is similar to that presented for the meson model\cite{Tang}.
Since the pion field vanishes in the mean-field approximation, 
the details of its construction are not presented here, 
nor is an explicit form for these terms given.
The pion will first enter when considering two-loop contributions
to the energy.

Beyond a leading-order transformation from a meson lagrangian
to a point-coupling lagrangian,
the full expressions for the mean meson fields involve 
infinite series in powers of the various bilinears of the nucleon field.
A precise transformation from the meson effective lagrangian
at the mean-field level would
therefore lead to the presence of higher-order terms in a 
point-coupling model that are treated here as negligible.
In practice, because of delicate cancellations,
a simple truncation of these terms leads to a poor
fit with experimental data, but if the parameters are allowed
to readjust slightly 
the new fit to the data is quite good.  

This is not unlike the 
truncation of higher-order terms within the mean-field meson model:  the 
inclusion of fifth-order terms (see Eq.~(54) of Ref.~\cite{Tang})
into the lagrangian improves a fit to the data only marginally, but
if the fifth-order terms are then simply truncated from the lagrangian,
their effects must be absorbed into an adjustment of the 
lower-order coefficients if a good fit is to remain intact.
Thus, the correspondence between the point-coupling model used here and
the mean-field meson model of Ref.~\cite{Tang} is {\em not exact}
due to truncations.
Relations of the leading-order point-coupling coefficients here to
the corresponding mean-field meson model coefficients
that would result from a full transformation  are 
listed below
in  Tables~\ref{tab:coeffsFZ},  \ref{tab:coeffsVZ}, \ref{tab:coeffsFA},
and \ref{tab:coeffsVA}.

In the meson model of Furnstahl et~al.\cite{Tang}, the vector and $\rho$-meson
masses are  fixed at their vacuum values of $782\,$MeV and
$770\,$MeV, respectively.  
The correspondences of these masses to the point-coupling parameters
are given by
\begin{eqnarray}
     m_v^2 = \wt\zeta_2/\wt\zeta_d \ ,
          \qquad m_\rho^2 &=& \wt\xi_2/\wt\xi_d \ .\label{eq:masses}
\end{eqnarray}
The vector meson and $\rho$-meson masses can therefore 
be ``held fixed'' within
the point-coupling model by varying only
$\wt\zeta_2$ and $\wt\xi_2$ 
and determining $\wt\zeta_d$ and ${\wt\xi}_d$
through these relations.
These relations were not imposed in the point-coupling
model studied by Hoch et~al.\cite{Madland}.
Here we perform fits to experimental data both with these
combinations fixed and 
allowing the four parameters to 
vary independently.

\section{Single-Nucleon Properties}

The values of the photon coupling parameters 
$\lambda_n$, $\lambda_p$, $\wt\beta_s$, $\wt\beta_v$ and 
$\wt\delta_{\rm S}+\wt\delta_{\rm V}$
are all determined from single-nucleon electromagnetic
form factors in the vacuum, calculated at tree level.
The anomalous magnetic moments of the nucleon are fixed
at $\lambda_p = 1.793$ and $\lambda_n = -1.913$\cite{Tang}.
The other three parameters are related to the isoscalar and isovector charge
form factors, which are given here for spacelike momenta $Q^2 \equiv -q^2$ by
\begin{eqnarray}
F_1^s(Q^2) = {1 \over 2} - {\wt\beta_s\over 2M^2}Q^2 
         - {\wt\delta_s\over M^4}Q^4 + \cdots\ ;
\label{eq:scalarform}\\
F_1^v(Q^2) = {1 \over 2} - {\wt\beta_v\over 2M^2}Q^2 
      - {\wt\delta_v\over M^4}Q^4 +\cdots\ .
\label{eq:vectorform}
\end{eqnarray}
The coefficients of the second-order terms are proportional
to  the corresponding
mean-square charge radii, and their values are fixed
from the experimental values \cite{BROWN}:
\begin{eqnarray}
\wt\beta_s &=& M^2\langle r^2 \rangle_{s1}/6  \approx
	 M^2(0.79\,{\rm fm})^2/6 =  2.36\ ,\\
\wt\beta_v &=& M^2 \langle r^2 \rangle_{v1}/6  \approx
	 M^2(0.79\,{\rm fm})^2/6 =  2.36\ .
\end{eqnarray}
These coefficients combine the direct coupling and
VMD contributions that together  determine the charge
radii in the model of Ref.~\cite{Tang}.
Unlike that model, there is no density dependence here from
the effective masses of the vector mesons in medium.
Since the difference between the point
proton density and the full charge density is not very important
in determining the self-consistent wavefunctions and energy levels,
the point-coupling electromagnetic contributions produce 
similar results to the conventional convolution procedure \cite{Serot},
in which empirical single-nucleon form factors are folded with point
nucleon densities after self-consistency is reached.

\begin{figure}
\setlength{\epsfxsize}{4.0in}
\centerline{\epsffile{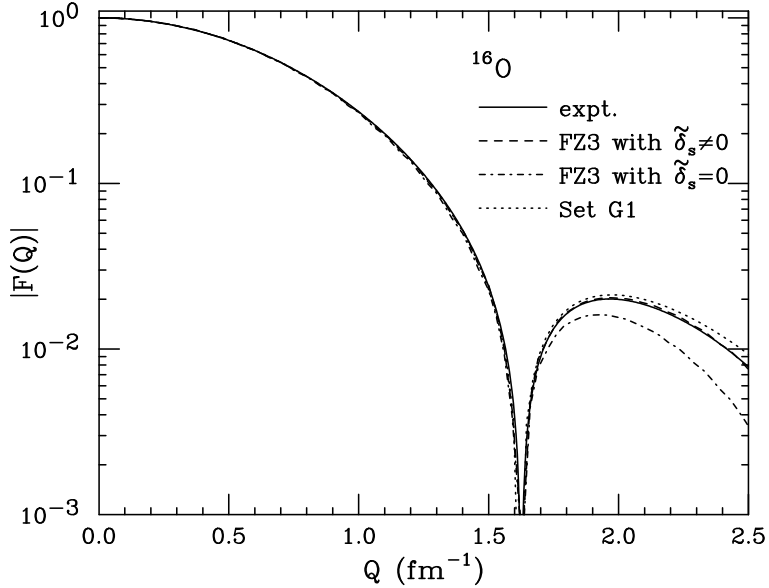}}
\caption{Charge form factor of $\null^{16}$O. The solid line
is taken from Ref.~\protect\cite{deVries}.  Form factors are shown for set
FZ3 of the point-coupling model with and without fourth-order momentum
corrections.  Also, set G1 from the meson model 
\protect\cite{Tang} is shown.}
\label{fig:FCH}
\end{figure}

Nuclear structure observables do not depend
strongly on the values of the fourth-order coefficients
$\wt\delta_s$ and $\wt\delta_v$.
Based on the criteria used here, the quality of fits are not
affected by the inclusion of these parameters and their associated
terms.  They do play a key role in  
the (momentum-space) charge form-factor of each nucleus
for higher values of $Q^2$, however.  The d.m.s.\ charge radius,
which depends on the position of the first zero of the charge
form-factor $Q_0^{(1)}$\cite{Tang}:
\begin{eqnarray}
R_{\rm dms} \equiv 4.493/Q_0^{(1)}\ ,
\end{eqnarray}
is used as one criterion for optimization, and
the parameter sets with good values of $\chi^2$ do reproduce the
experimental result for the value of $Q_0^{(1)}$.
However, 
if the fourth-order corrections are omitted,  for oxygen
there is a significant
deviation  from  experiment beyond this value of momentum.
In particular, the second maximum falls short
of the predicted value (see Fig.~\ref{fig:FCH}).
This is in contrast to the meson model, where
vector-meson dominance  yields the 
necessary momentum dependence to produce accurate results
for the second maximum without including it as
one of the criteria for optimization.

Fourth-order corrections in momentum are therefore included
in the model through the parameters $\wt\delta_s$ and $\wt\delta_v$.
A determination of the values of these parameters through vacuum
properties is sufficient to reproduce the second maximum
(see Fig.~\ref{fig:FCH}).
In fact, due to the small magnitude of the neutron charge
density in comparison to the proton,
only the isospin $+1/2$ components of the additional terms
are needed and $\wt\delta_s$ and $\wt\delta_v$ are therefore
set equal.

We use
a dipole fit to the proton form factor to determine the value of 
$\wt\delta_s = \wt\delta_v$.
(Any alternative parameterization of the form factor could be used 
instead; our results are not sensitive to the details because
it is a low-momentum expansion.)
The proton form factor, $G_E^p(Q^2)$ is known to be fitted well by a dipole 
form\cite{BHADURI}:
\begin{eqnarray}
   G_E^p(Q^2) = \left(1+ \alpha{ Q^2\over M^2}\right)^{-2}
     = 1 - 2 \alpha {Q^2\over M^2} + 3\alpha^2 {Q^4\over M^4} +\cdots\ .
               \label{eq:dipoleform}
\end{eqnarray}
This form factor is related to the isoscalar and isovector form factors by the
relation
\begin{eqnarray}
    G_E^p(Q^2) = F^p_1(Q^2) - {Q^2\over 4M^2}F^p_2(Q^2)\ ,\label{eq:chargeform}
\end{eqnarray}
where 
\begin{eqnarray}
   F_i^p(Q^2) \equiv F_i^s(Q^2) + F_i^v(Q^2)\ .
\end{eqnarray}
Using Eqs.~(\ref{eq:scalarform}) and (\ref{eq:vectorform}) and the
fact that $\wt\delta_s = \wt\delta_v$,
the proton form factor in Eq.~(\ref{eq:chargeform}) can be expanded  as
\begin{eqnarray}
G_E^p(Q^2) = 1 -{2\wt\beta_s+2\wt\beta_v+\lambda_p\over 4M^2}Q^2
  -{2\wt\delta_s\over M^4}Q^4 +\cdots\ .\label{eq:proton}
\end{eqnarray}
Comparison with Eq.(\ref{eq:dipoleform}) yields
\begin{eqnarray}
&\alpha = {\textstyle 2\wt\beta_s+2\wt\beta_v+\lambda_p\over 
         \mathstrut\textstyle 8} 
      = 1.39\ ;&\\
&2\wt\delta_s = -3\alpha^2 =  -5.80\ .&
\end{eqnarray} 

In principle, higher-order momentum corrections to the
anomalous form factors $F_2^{s,v}$, which are not considered
here, would lead to an additional $Q^4$ dependence to 
the proton form-factor.  Our sole interest is in determining
the {\em overall} coefficient appearing at fourth-order in 
Eq.~(\ref{eq:proton}), however, and such a contribution could be re-absorbed
into a redefinition of $\wt\delta_s$.

\section{The Mean-Field Solutions}

We work at one-baryon-loop order in this paper, which is equivalent
to the Dirac-Hartree approximation \cite{Serot}.  This should be sufficient
to test the consistency of naturalness through
NDA and the truncation of the lagrangian.
As discussed in Ref.~\cite{FURNSTAHL96}, a Hartree calculation can be viewed as
equivalent to a density-functional approach, in which higher-order
many-body corrections are treated approximately.
Such corrections have not yet been explored in a relativistic point-coupling
model.
In future work, the stability of the Hartree results will be tested
in a two-loop calculation.

From the lagrangian, one can derive the Dirac equation and the 
energy functional for nuclei conventionally,
keeping in mind that there are time derivatives acting on $\psi^\dagger$
as well as $\psi$.
(See Ref.~\cite{Madland} for details.)
Because of its large mass,
loop integrals with the nucleon include dynamics from distance scales
that are much shorter than the scale set by the valence-nucleon momenta,
which are limited by the Fermi momentum $k_{\rm F}$.
These short-range effects are included implicitly in the coefficients
of the lagrangian and energy functional.  Formally, one can include
counterterms to remove these loop effects to all orders, which is always
possible, since all such terms are already contained in the effective
lagrangian.  

The  single-particle Dirac hamiltonian for spherically symmetric
nuclei takes a form similar
to that of Ref.~\cite{Tang}:
\begin{eqnarray}
h({\bf x}) &=& -i\bbox{\alpha\cdot}\grad + W(r) +{1\over 2}\tau_3 R(r) 
	+ \beta (M - \Phi(r)) 
	+ i\beta\bbox{\alpha\cdot T}(r)
	+{i\over 2}\beta\bbox{\alpha\cdot T}_3(r)\tau_3\nonumber\\
&&	+{1\over2}(1+\tau_3)A(r)
	+{1\over 2M^2}(\beta_s + \beta_v\tau_3)\grad^2 A(r)
	-{1\over M^4}(\delta_s + \delta_v\tau_3)(\grad^2)^2A(r)
	\nonumber\\
&&	-{i\over 4M}[\lambda_p(1+\tau_3) + \lambda_n(1-\tau_3)]
        \beta\bbox{\alpha\cdot}\grad A(r)
	\ ,
\end{eqnarray}
where the potentials $\Phi$, $W$, $R$, $A$, $\bbox{T}$, and $\bbox{T}_3$ 
are defined by
\begin{eqnarray}
\Phi &\equiv& -\Lambda\{2\wt\kappa_2\scalarN 
	+ 2\wt\kappa_d{\wt\grad}^2\scalarN
	+3\wt\kappa_3\scalarN^2
	+4\wt\kappa_4\scalarN^3
	+\wt\eta_1(\vectorN^0)^2
	+2\wt\eta_2(\vectorN^0)^2\scalarN
	+\wt\eta_\rho(\isovectorVectorN^0)^2\nonumber\\
&&	\null
        +\wt\alpha_1({\wt\grad}\scalarN)^2 
	+2\wt\alpha_1\scalarN{\wt\grad}^2\scalarN
	-\wt\alpha_2({\wt\grad}\vectorN^0)^2 \}\ ,\\
W    &\equiv& \Lambda\{2\wt\zeta_2(\vectorN^0)
	+2\wt\zeta_d{\wt\grad}^2(\vectorN^0)
	+4\wt\zeta_4(\vectorN^0)^3
	+2\wt\eta_1\vectorN^0\scalarN
	+2\wt\eta_2\vectorN^0\scalarN^2\nonumber\\
&&	\null 
        +2\wt\alpha_2 ( {\wt\grad}\scalarN\bbox{\cdot}
		{\wt\grad}\vectorN^0
	          + \scalarN{\wt\grad}^2\vectorN^0 )
	+\wt f_v{\wt\grad}\bbox{\cdot\tensorN}\}\ ,\\
R    &\equiv& \Lambda\{2\wt\xi_2\isovectorVectorN^0
	+2\wt\xi_d{\wt\grad}^2\isovectorVectorN^0
	+2\wt\eta_\rho\scalarN\isovectorVectorN^0
	+\wt f_\rho{\wt\grad}\bbox{\cdot\isovectorTensorN}\}\ ,\\
A    &\equiv& eA_0\ ;\\
\bbox{T}  &\equiv& -\Lambda \wt f_v{\wt\grad}\vectorN^0\ ,\\
\bbox{T}_3 &\equiv& 
 -\Lambda\wt f_\rho{\wt\grad}\isovectorVectorN^0\ ,\\
\bbox{\tensorN}_i &\equiv& \tensorN_{0i}\ ,\\
\bbox{\isovectorTensorN}_i &\equiv& \isovectorTensorN{}_{0i}\ .
\end{eqnarray}
The Dirac equation with eigenvalues $E_\alpha$ and 
eigenfunctions $N_\alpha({\bf x})$ is
\begin{equation}
  h N_\alpha({\bf x}) = E_\alpha N_\alpha({\bf x}) \ ,
   \qquad\qquad  
   \int\!d^3x\ N^\dagger_\alpha({\bf x})N_\alpha({\bf x}) = 1 \ .
\end{equation}
Following the conventions of Ref.~\cite{Serot} (also used in Ref.~\cite{Tang}),
the eigenfunctions  for  spherically symmetric nuclei are written 
in terms of spin spherical harmonics, $\Phi_{\kappa m}$:
\begin{eqnarray}
N_\alpha &=& N_{n\kappa m t} = \left(\matrix{
{\textstyle i\over \textstyle r}G_a(r)\Phi_{\kappa m}\cr
-{\textstyle 1\over \textstyle r}F_a(r)\Phi_{-\kappa m}
}\right)\zeta_t\ ,
\end{eqnarray}
where $t=1/2$ for protons and $t=-1/2$ for neutrons.
The equation for $N_\alpha$ then reduces to a set of 
coupled equations for $G_a$ and $F_a$:
\begin{eqnarray}
&&\left({d\over dr} + {\kappa\over r}\right)G_a(r)
-[E_\alpha - U_1(r) + U_2(r)]F_a(r) - U_3(r)G_a(r) = 0\ ,\\
&&\left({d\over dr} - {\kappa\over r}\right)
+[E_\alpha - U_1(r) - U_2(r)]G_a(r) + U_3(r)F_a(r) = 0\ ,
\end{eqnarray}
where we have defined the single-particle potentials by
\begin{eqnarray}
U_1(r) &=& W(r) + t_a R(r) + (t_a +1/2)A(r) 
	+{1\over 2M^2}(\wt\beta_s + 2t_a\wt\beta_v)\grad^2 A(r)\nonumber \\
&&\quad
	-{1\over M^4}(\wt\delta_s + 2t_a\wt\delta_v)(\grad^2)^2A(r)\ ,\\
U_2(r) &=& M - \Phi(r)\ ,\\
U_3(r) &=& -\bbox{T}(r)\cdot \hat r - t_a\bbox{T_3}(r)\cdot\hat r
	+{1\over 2M}[(\lambda_p+\lambda_n)/2
		+ t_a(\lambda_p-\lambda_n)] {dA\over dr} \ .
\end{eqnarray}

The mean-field densities can be
expressed in terms of the radial wave functions $G$ and $F$ in the 
same manner as Refs.~\cite{Tang} and \cite{Serot}:
\begin{eqnarray}
\scalarN(r)& = &
       {1 \over f_\pi^2\Lambda}
	\sum_{a}^{\rm occ}{2j_a+1\over 4\pi  r^2}
	\Bigl(G_a^2(r)  - F_a^2(r)\Bigr)\ ,\\
\vectorN^0(r) & = &
       {1 \over f_\pi^2\Lambda}
	\sum_{a}^{\rm occ}{2j_a+1\over 4\pi r^2}
	\Bigl(G_a^2(r)  + F_a^2(r)\Bigr)\ ,\\
\isovectorVectorN^0(r)& = &
       {1 \over f_\pi^2\Lambda}
	\sum_{a}^{\rm occ}{2j_a+1\over 4\pi r^2}
	(t_a)\Bigl(G_a^2(r)  + F_a^2(r)\Bigr)\ , \\
\bbox{\tensorN}(r)\cdot\hat r & = &
       {1 \over f_\pi^2\Lambda}
	\sum_{a}^{\rm occ}{2j_a+1\over 4\pi r^2}\,
	2G_a(r)F_a(r)\ ,\\
\bbox{\isovectorTensorN}(r)\cdot\hat r & = &
       {1 \over f_\pi^2\Lambda}
	\sum_{a}^{\rm occ}{2j_a+1\over 4\pi r^2}
	(t_a)\,2G_a(r)F_a(r)\ .
\end{eqnarray}
The summation superscript ``occ'' means that the sum runs only over
occupied (valence) states in the Fermi sea.
The mean-field equation for the photon field is given by
\begin{equation}
-\grad^2 A = e^2 \rho_{\rm chg}(r)\ , 
\end{equation}
where the charge density is
\begin{eqnarray}
\rho_{\rm chg}(r) 
      &\equiv& \sum_{\alpha}^{\rm occ} \Bigl[
      \overline N_\alpha{1\over 2}(1+\tau_3)\gamma_0 N_\alpha
+{1\over 2M^2}\grad^2\overline N_\alpha(\wt\beta_s + \wt\beta_v\tau_3)
 \gamma_0 N_\alpha
\nonumber\\
& & \null
      -{1\over M^4}(\grad^2)^2\overline N_\alpha(\wt\delta_s + 
        \wt\delta_v\tau_3)\gamma_0 N_\alpha
    +{i\over 2M}\grad\cdot\overline N_\alpha
         \lambda\beta\bbox{\alpha} N_\alpha \Bigr]\ .
\end{eqnarray}

The Hartree equations are solved by an iterative procedure similar
to that used for meson models \cite{Serot}.
The procedure is simplified over that of the meson model in that
only one non-linear differential equation for a meson field
(the Coulomb field $A$) is solved; the input potentials
for each iteration
of the Dirac equation are then just evaluated in terms of the densities
from the previous iteration.
See Ref.~\cite{Madland} for more details.

An expression for the energy is given by the first diagonal
element $T_{00}$ of the mean-field stress-energy tensor:
\def\lagn{{\cal L}}
\begin{eqnarray}
E &=& \int\!d^3x\, T_{00}  \nonumber\\ 
&=&\int\!d^3x\,\sum_{\alpha}^{\rm occ}\,\overline N_\alpha(
-i\beta\bbox{\alpha\cdot}\grad + M)N_\alpha 
+ f_\pi^2\Lambda^2\int\!d^3x\,\bigg\{
\wt\kappa_2\scalarN^2
 -\wt\kappa_d[ (\delN^0\scalarN)^2
             + ({\wt\grad}\scalarN)^2]  \nonumber\\
&&\qquad\qquad\null
+\wt\kappa_3\scalarN^3
+\wt\kappa_4\scalarN^4
+\wt\eta_1(\vectorN^0)^2\scalarN
+\wt\eta_2(\vectorN^0)^2\scalarN^2\nonumber\\
&&\qquad\qquad\null+\wt\zeta_2(\vectorN^0)^2
-\wt\zeta_d [ (\delN^0\vectorN^0)^2
                 + ({\wt\grad}\vectorN^0)^2 ]
+\wt\zeta_4(\vectorN^0)^2(\vectorN^0)^2\nonumber\\
&&\qquad\qquad\null+\wt\xi_2(\isovectorVectorN^\mu)^2
-{\wt\xi}_d  [(\delN^0\isovectorVectorN^\mu)^2
               + ({\wt\grad}\isovectorVectorN^\mu)^2 ]
+\wt\eta_\rho(\isovectorVectorN^\mu)^2\scalarN\nonumber\\
&&\qquad\qquad\null
-\wt\alpha_1 \scalarN [ (\delN_0\scalarN)^2
                        + ({\wt\grad}\scalarN)^2]
-\wt\alpha_2 \scalarN [ (\delN_0\vectorN^0)^2
                        + ({\wt\grad}\vectorN^0)^2]
\nonumber\\
&&\qquad\qquad\null
  +\wt f_v{\wt\grad}\vectorN^0\cdot\bbox{\tensorN}
  +\wt f_\rho
	{\wt\grad}\isovectorVectorN^0\cdot\bbox{\isovectorTensorN}
  \bigg\} \nonumber\\
&&\qquad\qquad\null + \mbox{electromagnetic terms}\ .
\end{eqnarray}
Ground-state densities are time-independent, so all time derivatives
in $E$ vanish.
The Dirac equation can be used to replace the first two terms
with a new expression involving a sum over the energies of the occupied
states; after integrating by parts and using the 
equation for the electromagnetic potential, 
the expression becomes
\begin{eqnarray}
E &=&\sum_{\alpha}^{\rm occ}E_\alpha -\int\!d^3x\,
f_\pi^2\Lambda^2\bigg\{
\wt\kappa_2\scalarN^2
+\wt\kappa_d\scalarN{\wt\grad}^2\scalarN
+2\wt\kappa_3\scalarN^3
+3\wt\kappa_4\scalarN^4\nonumber\\
&&\qquad\qquad\null+\wt\zeta_2(\vectorN^0)^2
+\wt\zeta_d\vectorN{\wt\grad}^2\vectorN^0
+3\wt\zeta_4(\vectorN^0)^4\nonumber\\
&&\qquad\qquad\null+2\wt\eta_1(\vectorN^0)^2\scalarN
+3\wt\eta_2(\vectorN^0)^2\scalarN^2\nonumber\\
&&\qquad\qquad\null+\wt\xi_2(\isovectorVectorN^0)^2
+{\wt\xi}_d\isovectorVectorN^0{\wt\grad}^2\isovectorVectorN^0
+2\wt\eta_\rho(\isovectorVectorN^0)^2\scalarN\nonumber\\
&&\qquad\qquad\null
- 2\wt\alpha_1 \scalarN ({\wt\grad}\scalarN)^2
-2\wt\alpha_2 \scalarN ({\wt\grad}\vectorN^0)^2	
\nonumber\\
&&\qquad\qquad\null
+\wt f_\rho
	\isovectorVectorN^0{\wt\grad}\cdot\bbox{\isovectorTensorN}
+\wt f_v\vectorN^0{\wt\grad}\cdot\bbox{\tensorN}\bigg\}
- {1\over 2}\int\!d^3x\, A \rho_{\rm chg}\ .
\end{eqnarray}
A center-of-mass correction 
for the energy is incorporated
as in Ref.~\cite{Tang} but the charge radius is not corrected
(including the latter does not change our conclusions).

\section{Optimization}

All parameters of the model other than those fixed by single-nucleon
properties in the vacuum (see Sect.~III)
are treated as free variables
and an optimum fit to experimental nuclear data is sought.
The same observables and weights are used as in
Ref.~\cite{Tang}. 
This optimization faces serious challenges.
The existence of delicate cancellations at lowest order in the
density expansion necessitates 
calculations to higher order in the density 
to obtain an adequate description of nuclei.
This increases the size of the 
parameter set to optimize and since
 the observables themselves are highly correlated, there
is a problem with underdetermination. 
The delicate cancellations 
also result in an extreme sensitivity to the lower-order parameters,
precluding large steps through the parameter space if the 
Hartree iterations are to remain stable;
the optimization program must therefore
be efficient at navigating narrow valleys.

Nonetheless, the optimization procedure
has been significantly improved over that used in obtaining the results
in Ref.~\cite{Tang}.  
This has been achieved in part by 
exploiting the relatively small difference between scalar and vector
densities at ordinary nuclear densities.
We can
rewrite the lagrangian as an expansion in
the bilinears $\wt\rho_+$ and $\wt\rho_-$ defined 
by\footnote{An analogous procedure
can be performed in the meson model \cite{Tang} 
by rewriting the lagrangian in terms
of the sum and difference of the scalar and (zero component of) the
vector fields, $(m_s^2\Phi/g_s^2 \pm m_v^2 W/g_v^2)/2$.}
\begin{eqnarray}
\wt\rho_+ &\equiv& 
   \sum_{\alpha}^{{\rm occ}}
{\overline N_\alpha(1+\gamma_0)N_\alpha\over 2f_\pi^2\Lambda}
= {1\over 2}(\scalarN + \vectorN^0)
\ ;\\
\wt\rho_- &\equiv& 
   \sum_{\alpha}^{{\rm occ}}
     {\overline N_\alpha(1-\gamma_0)N_\alpha\over 2f_\pi^2\Lambda}
= {1\over 2}(\scalarN - \vectorN^0)\ ,
\end{eqnarray}
where the sum is over occupied states $N_\alpha$.
This separation is reminiscent of the heavy baryon formalism (HBF)\cite{HBF1,HBF2}
in the vacuum,
with the spin matrix $(1-\gamma_0)/2$ acting to project out
the negative energy states in the vacuum (at least to leading
order in an inverse nucleon mass expansion).

Here we simply observe that
a useful hierarchy for optimization is given  in terms of
``optimal parameters,'' which are
listed in  Table~\ref{tab:optcoeffs}.
These parameters are linear combinations of the couplings in the
lagrangian.
The hierarchy is based on the observation that the
difference $\wt\rho_-$
between scalar and vector densities is small  
and empirically scales like
$\wt\rho_+^{8/3}$ near equilibrium density.%
\footnote{At low density $\wt\rho_-$ scales like $\wt\rho_+^{5/3}$,
but this density dependence is independent of the parameters.} 
We can classify terms according to a power of $\wt\rho_+$,
which is roughly $1/4$ to $1/7$.
The label in the first column of Table~\ref{tab:optcoeffs} indicates how
different parameter sets are organized (see below);
it does not correspond 
precisely to a systematic expansion in $\wt\rho_+$.
Performing the optimization in terms of the $\wt\Omega_i$ parameters
in Table~\ref{tab:optcoeffs} is more efficient; 
we stress, however, that the energy functional has not been changed.

\begin{table}[tb]
\caption{Hierarchy of ``optimal'' terms (see text).  
Terms are presented in order
of importance from top to bottom.  Derivative terms are treated separately.}
\begin{tabular}{c|l|c|c}
\multicolumn{2}{c|}{Without derivatives}
&\multicolumn{2}{c} {Terms in lagrangian}\\ \hline
set & optimal parameter  & new notation  & covariant  \\ \hline
\hline
      $\begin{array}{c} P_{0} \\ P_{0} 
       \end{array}$     &  
      $\begin{array}{l} \opt{1} = \wt\kappa_2+\wt\zeta_2 \\ 
                        \opt{2} = \wt\kappa_2-\wt\zeta_2 
       \end{array}$     &
      $\opt{1}(\wt\rho_+^2+\wt\rho_-^2) +
                     2 \opt{2}(\wt\rho_+\wt\rho_-)$     & 
           $\wt\kappa_2\scalarN^2 + \wt\zeta_2(\vectorN^0)^2$
    \\ \hline
      $\begin{array}{c} P_{1} \\ P_{2} 
       \end{array}$     &  
      $\begin{array}{l} \opt{3} =  \wt\kappa_3 + \wt\eta_1 \\ 
                        \opt{4} =  \wt\kappa_3 - \wt\eta_1/3 
       \end{array}$     &
      $\opt{3}(\wt\rho_+^3 + \wt\rho_-^3)
               + 3\opt{4}(\wt\rho_+^2\wt\rho_- +\wt\rho_+\wt\rho_-^2)$  & 
      $\wt\kappa_3\scalarN^3 +\wt\eta_1(\vectorN^0)^2\scalarN$
    \\ \hline
      $\begin{array}{c} P_{3} \\ P_{4} \\ P_{5} 
       \end{array}$     &  
      $\begin{array}{l} \opt{5} = \wt\kappa_4 +\wt\zeta_4 + \wt\eta_2 \\ 
                        \opt{6} = \wt\kappa_4 - \wt\zeta_4 \\
                        \opt{7} = \wt\kappa_4 + \wt\zeta_4 - \wt\eta_2/3 
       \end{array}$     &
      $\begin{array}{c}
             \opt{5}(\wt\rho_+^4 +\wt\rho_-^4)+
                4 \opt{6}(\wt\rho_+^3\wt\rho_- +\wt\rho_+\wt\rho_-^3) \\
             \null + 6\opt{7}(\wt\rho_+^2\wt\rho_-^2)
       \end{array}$  &
      $\wt\kappa_4\scalarN^4 + \wt\zeta_4(\vectorN^0)^4
              +\wt\eta_2\scalarN^2(\vectorN^0)^2$
    \\ \hline\hline
\multicolumn{2}{c|}{With derivatives}&
\multicolumn{2}{c}{}\\ \hline
      $\begin{array}{c} P_{0} \\ P_{0} 
       \end{array}$     &  
      $\begin{array}{l} \optder{1} = \kappa_d+\zeta_d \\ 
                        \optder{2} = \kappa_d-\zeta_d 
       \end{array}$     &
      $\begin{array}{c}
             \optder{1}[(\delN_\mu\wt\rho_+)^2
                        +(\delN_\mu\wt\rho_-)^2] \\
             \null+ 2\optder{2}\delN_\mu\wt\rho_+\delN^\mu\wt\rho_-
       \end{array}$  &
      $\wt\kappa_d(\delN_\mu\wt\scalar)^2 
                 + \wt\zeta_s(\delN_\mu\vectorN{}_\nu)^2$
    \\ \hline
      $\begin{array}{c} P_{2} \\ P_{2} 
       \end{array}$     &  
      $\begin{array}{l} \optalpha{1} = \wt\alpha_1+\wt\alpha_2 \\ 
                        \optalpha{2} = \wt\alpha_1-\wt\alpha_2 
       \end{array}$     &
      $\begin{array}{c}
             \optalpha{1}[(\delN_\mu\wt\rho_+)^2 
               +(\delN_\mu\wt\rho_-)^2]  (\wt\rho_+ + \wt\rho_-) \\
             \null + 2\optalpha{2}\wt\del_\mu\wt\rho_+\wt\del^\mu\wt\rho_-
                     (\wt\rho_+ + \wt\rho_-)
       \end{array}$  &
      $\wt\alpha_1(\delN_\mu\scalarN)^2\scalarN 
              + \wt\alpha_2(\delN_\mu\vectorN{}_\nu)^2\scalarN$
\end{tabular}
\label{tab:optcoeffs}
\end{table}

Improvements were also made to the optimization procedure
by keeping careful track of the necessary precision of the parameters.
The calculation of $\chi^2$ 
(treated as a function of the optimal parameters)
is much more sensitive to small changes in 
the lower-order parameters than changes of the same size in the higher-order
parameters.  The difference is severe enough to create problems
for the minimization software. 
To remedy this, each parameter was expressed in the form
\begin{eqnarray}
\Omega_i = \Omega_i^{\rm fixed} + 10^{-\lambda_i}\delta\Omega_i\ ,
\end{eqnarray}
where $\Omega_i^{\rm fixed}$ and $\lambda_i>0$
are determined by the initial values of the parameters and are held
fixed during optimization.  The parameters are then varied
by changing the value of $\delta\Omega_i$.  Lower-order parameters have a 
larger value of $\lambda_i$ so that large changes in $\delta\Omega_i$ 
actually correspond to small changes in the true parameter $\Omega_i$.
The values of $\Omega_i$ and $\lambda_i$ were chosen
to yield roughly similar curvatures of the $\chi^2$--function
for variations of each $\delta\Omega_i$
and therefore yield a similar sensitivity of $\chi^{2}$ 
to each parameter.
Given the delicate dependence on the lower-order parameters, 
an advantage was also gained over the optimization procedure
used in the meson model by creating a routine {\it outside
the minimization package\/} to calculate the gradient
of the $\chi^2$--function (with respect to the optimal parameters) 
needed by the optimizer.

\section{Results and Discussion}

The model is fit to  
experimental data for four different categories of
parameter sets, each at different levels of truncation.
The parameter sets are categorized according to three alphanumeric labels.
Optimizations in which the fixed-mass relations from the meson
model are imposed on the parameters $\zeta_d$ and $\xi_d$
are labeled with an initial letter of ``F'',
and those in which $\zeta_d$ and $\xi_d$ are freely varied are labeled
with an initial letter of ``V''.
Optimizations in which  the parameters 
$\wt\alpha_1$ and $\wt\alpha_2$ are fixed at 
zero are labeled with a second letter of ``Z'', and those
in which they are freely varied are labeled with a second letter of ``A''.
The nomenclature for each optimized parameter set
also includes a number reflecting
the level of truncation (see Table~\ref{tab:optcoeffs}):
\begin{center}
\begin{tabular}{l}
 parameter set of terms kept\\
$ P_0 \equiv \{\opt{1},\opt{2},
	\optder{1},\optder{2}, \wt\xi_2, \wt\xi_{\rm d}\}$\\
$P_1 \equiv P_0 \cup\{\opt{3},\wt f_\rho, \wt f_{\rm v}\}$\\
$P_2 \equiv P_1 \cup\{\opt{4},\wt\eta_\rho\}$\\
$P_3 \equiv P_2\cup\{\opt{5}\}$\\
$P_4 \equiv P_3\cup\{\opt{6}\}$\\
\end{tabular}
\end{center}
We stress that the number denoting the level of truncation
{\em does not} correspond with an expansion to that order
in the scalar and vector densities, 
but instead to an organization according to the 
optimal parameters.
However, we can identify set~$P_{0}$ with a truncation at
second order in the densities,
set~$P_{2}$ with third order in the densities, and set~$P_{4}$ with
fourth order in the densities.
Sets FA$i$ and VA$i$ also include
the parameters $\wt\alpha_1$ and $\wt\alpha_2$; 
only values of $i \ge 2$ are considered in
these categories.

\begin{table}[tbp]
\caption{Parameter sets from fits to finite nuclei.
 Fixed $m_v$, $m_\rho$ and $\wt\alpha_1=\wt\alpha_2=0$.}
\begin{tabular}{ccrrrrr}
\ \ Point-Coupling\ \ &\ \  Meson \ \ & \ \ \ FZ$0$\ \ \  
&\ \ \  FZ$1$\ \ \  &  \ \ \  FZ$2$\ \ \  & \ \ \ FZ$3$\ \ \ &\ \ \ FZ$4$\ \ \ 
 	\\
\hline
$\wt\kappa_2$& $-g_s^2\fpi^{2}/(2m_s^2)$           
&$-1.930$ &$-1.229$  &$-1.566$  &$-0.963$ & $-1.042$\\
$\wt\zeta_2$& $g_v^2\fpi^{2}/(2m_v^2)$             
& $1.490$ & $0.760$  & $1.079$  & $0.460$ &  $0.538$\\
$\wt\xi_2$ & $g_\rho^2\fpi^{2}/(2m_\rho^2)$        
& $0.679$ & $0.524$  & $0.664$  & $1.014$ &  $0.992$\\
$\wt\eta_1$&                    
&         & $0.521$  & $-1.021$ & $1.615$ & $0.658$\\
$\wt\eta_2$&    
&         &          &          & $-1.178$ &  $-1.207$\\
$\wt\kappa_3$&                 
&         & $0.174$  & $1.784$  & $-0.368$ &  $0.608$\\
$\wt\kappa_4$&  
&         &          &          & $-0.196$ & $-1.530$\\
$\wt\zeta_4$& 
&         &          &          & $-0.196$ &  $1.128$\\
$\wt\eta_\rho$&              
&         &          & $-1.142$ & $-3.400$ & $-3.245$\\
$\wt f_v$&                                
&         & $0.411$  & $0.434$  & $0.667$ & $0.685$\\
$\wt f_\rho$           &                              
&         & $2.503$  & $1.904$  & $1.659$ & $1.575$\\
$\wt\kappa_d$& $-g_s^2\fpi^{2}\Lambda^{2}/(2m_s^4)$           
&$-3.140$ & $-2.006$ & $-2.406$  & $-1.779$ & $-1.859$\\
$\wt\zeta_d$& $g_v^2\fpi^{2}\Lambda^{2}/(2m_v^4)$             
& $1.445$ & $0.737$  & $1.046$ & $0.446$ & $0.522$\\
${\wt\xi}_d$& $g_\rho^2\fpi^{2}\Lambda^{2}/(2m_\rho^4)$       
& $0.679$ & $0.524$  & $0.664$  & $1.014$ & $0.992$\\
\hline
$\wt\kappa_2/\wt\kappa_d$ & $m_s^2/\Lambda^{2}$ 
& $(0.784)^2$ & $(0.783)^2$  & $(0.807)^2$  &$(0.736)^2$ &  $(0.749)^2$\\
$-\wt\kappa_2^2\Lambda^2/(f_\pi^2\wt\kappa_d)$ 
& $g_s^2/(4\pi)$                               
& $1.009$ & $0.640$  & $0.866$  & $0.443$ &  $0.497$\\
$\wt\zeta_2\Lambda^2/(f_\pi^2\wt\zeta_d)$ & $g_v^2/(4\pi)^2$     
& $1.306$ & $0.666$  & $0.946$  & $0.404$ &  $0.471$\\
$\wt\xi_2^2\Lambda^2/(f_\pi^2{\wt\xi}_d )$ 
& $g_\rho^2/(4\pi)^2$                            
& $0.577$ & $0.445$  & $0.564$  & $0.862$ &  $0.843$\\
\hline
$\chi^2$ &   					  
& $1600$  & $65$  & $41$ & $37$ &  $37$\\
\end{tabular}
\label{tab:coeffsFZ}
\end{table}

\begin{table}[tbp]
\caption{Parameter sets from fits to finite nuclei.
Freely varied $m_v$, $m_\rho$, $\wt\alpha_1=\wt\alpha_2=0$.}
\begin{tabular}{ccrrrrr}
\ \ Point-Coupling\ \ &\ \  Meson \ \ & \ \ \ VZ$0$\ \ \  
&\ \ \  VZ$1$\ \ \  &\ \ \  VZ$2$\ \ \  
&\ \ \  VZ$3$ \ \ \  & \ \ \ VZ$4$\ \ \ \\ \hline
$\wt\kappa_2$& $-g_s^2\fpi^{2}/(2m_s^2)$           
&$-1.859$ &$-1.238$  &$-1.972$  & $-1.509$ & $-1.482$\\
$\wt\zeta_2$& $g_v^2\fpi^{2}/(2m_v^2)$             
& $1.427$ & $0.768$  & $1.482$  & $0.988$  & $0.961$\\
$\wt\xi_2$ & $g_\rho^2\fpi^{2}/(2m_\rho^2)$        
& $0.596$ & $0.535$  & $0.258$  & $0.345$  & $0.521$\\
$\wt\eta_1$&                    
&         & $0.526$  & $-1.906$ & $1.067$ & $0.055$\\
$\wt\eta_2$&    
&         &          &          &$-1.929$ & $-1.900$\\
$\wt\kappa_3$&                 
&         & $0.175$  & $2.509$  & $0.202$ & $1.278$\\
$\wt\kappa_4$&  
&        &           &          &$-0.321$ & $-2.105$\\
$\wt\zeta_4$&         
&        &           &          &$-0.321$ & $1.471$\\
$\wt\eta_\rho$&               
&        &           & $1.677$  &$1.133$ & $-0.220$\\
$\wt f_v$&                                
&       & $0.412$   & $0.027$  & $0.041$ & $0.189$\\
$\wt f_\rho$         &                              
&        & $2.836$&  $2.889$    & $3.780$ & $3.267$ \\ 
$\wt\kappa_d$& $-g_s^2\fpi^{2}\Lambda^{2}/(2m_s^4)$           
&$-0.732$&$-2.531$   & $-1.740$ &$-0.639$ & $-0.712$\\
$\wt\zeta_d$& $g_v^2\fpi^{2}\Lambda^{2}/(2m_v^4)$             
&$-0.739$& $1.238$   & $0.389$  &$-0.673$ & $-0.603$\\
${\wt\xi}_d$& $g_\rho^2\fpi^{2}\Lambda^{2}/(2m_\rho^4)$       
&$-3.813$& $1.074$   &$-1.578$  &$-3.060$ & $-2.410$\\ 
\hline
$\wt\kappa_2/\wt\kappa_d$ & $m_S^2/\Lambda^{2}$ 
& $(1.594)^2$ & $(0.699)^2$  & $(1.065)^2$  & $(1.537)^2$ &  $(1.443)^2$\\
$\wt\zeta_2/\wt\zeta_d$ & $m_v^2/\Lambda^{2}$ 
& $-1.931$ & $0.620$  & $3.811$  & $-1.468$ &  $-1.593$\\
$\wt\xi_2/\wt\xi_d$ & $m_\rho^2/\Lambda^{2}$ 
& $-0.156$ & $0.498$  & $-0.164$  & $-0.112$ &  $-0.216$\\
$-2\wt\kappa_2^2\Lambda^2/(f_\pi^2\wt\kappa_d)$ 
& $g_s^2/(16\pi^2)$                               
& $4.014$  & $0.515$   & $1.900$  & $3.031$&  $2.625$\\
$2\wt\zeta_2\Lambda^2/(f_\pi^2\wt\zeta_d)$ & $g_v^2/(16\pi^2)$    
& $-2.342$ & $0.404$   & $4.801$  &$-1.233$& $-1.300$\\
$2\wt\xi_2^2\Lambda^2/(f_\pi^2{\wt\xi}_d)$ 
& $g_\rho^2/(16\pi^2)$                            
& $-0.079$ & $0.226$  & $-0.036$ &$-0.033$ &$-0.096$\\
\hline
$\chi^2$ &   					  
& 1230  & 62  & $35$    & $27$ & $26$\\
\end{tabular}
\label{tab:coeffsVZ}
\end{table}

\begin{table}[tbp]
\caption{Parameter sets from fits to finite nuclei.
Fixed $m_v$, $m_\rho$. Nonzero $\wt\alpha_1$ and $\wt\alpha_2$.}
\begin{tabular}{ccrrr}
\ \ Point-Coupling\ \ &\ \  Meson \ \   
&\ \ \  FA$2$\ \ \  &\ \ \  FA$3$\ \ \  
&\ \ \  FA$4$ \ \ \  \\ \hline
$\wt\kappa_2$& $-g_s^2\fpi^{2}/(2m_s^2)$           
&$-2.065$  &$-1.306$  & $-1.298$ \\
$\wt\zeta_2$& $g_v^2\fpi^{2}/(2m_v^2)$             
& $1.563$  & $0.789$  & $0.782$  \\
$\wt\xi_2$ & $g_\rho^2\fpi^{2}/(2m_\rho^2)$        
& $0.466$  & $1.050$  & $1.045$  \\
$\wt\eta_1$&                    
&$-2.512$  & $1.007$ & $0.665$ \\
$\wt\eta_2$&   
&          & $-1.513$ &$-1.547$ \\
$\wt\kappa_3$&                 
& $3.200$  & $0.276$  &$0.625$ \\
$\wt\kappa_4$&  
&          & $-0.252$ &$-1.125$ \\
$\wt\zeta_4$ &         
&          & $-0.252$ &$0.609$  \\
$\wt\eta_\rho$&               
& $0.107$  & $-3.926$  &$-3.864$  \\
$\wt f_v$&                                
& $0.000$  & $0.315$  & $0.342$ \\ 
$\wt f_\rho$         &                              
& $2.449$  &  $1.857$ & $1.856$ \\ 
$\wt\kappa_d$& $-g_s^2\fpi^{2}\Lambda^{2}/(2m_s^4)$           
&$-3.117$  & $-2.190$ & $-2.136$ \\
$\wt\zeta_d$& $g_v^2\fpi^{2}\Lambda^{2}/(2m_v^4)$             
& $1.515$  & $0.765$  & $0.758$\\
${\wt\xi}_d$& $g_\rho^2\fpi^{2}\Lambda^{2}/(2m_\rho^4)$       
& $0.466$ & $1.050$  & $1.045$\\
$\wt\alpha_1$         &                              
& $16.42$  &  $11.138$ &$9.125$ \\ 
$\wt\alpha_2$         &                              
& $-13.12$ &  $-9.818$ &$-8.387$\\ \hline
$\wt\kappa_2/\wt\kappa_d$ & $m_S^2/\Lambda^2$ 
& $(0.814)^2$  & $(0.772)^2$  & $(0.779)^2$ \\ 
$-\wt\kappa_2^2/(8\pi^2\wt\kappa_d)$ 
& $g_s^2/(16\pi^2)$                               
& $1.162$  & $0.662$  &$0.670$ \\
$\wt\zeta_2/(8\pi^2\wt\zeta_d)$ & $g_v^2/(16\pi^2)$    
& $1.372$   & $0.692$  & $0.686$ \\
$\wt\xi_2^2/(8\pi^2{\wt\xi}_d)$ 
& $g_\rho^2/(16\pi^2)$                            
& $0.396$  & $0.893$  & $0.888$ \\ \hline
$\chi^2$ &   					  
& $36$  &  $33$  & $32$\\
\end{tabular}
\label{tab:coeffsFA}
\end{table}

\begin{table}[tbp]
\caption{Parameter sets from fits to finite nuclei.
Freely varied $m_v$, $m_\rho$, $\wt\alpha_1$ and $\wt\alpha_2$.}
\begin{tabular}{ccrrrr}
\ \ Point-Coupling\ \ &\ \  Meson \ \ 
&\ \ \  VA$2$\ \ \  &\ \ \  VA$3$\ \ \  
&\ \ \  VA$4$ \ \ \  \\ \hline
$\wt\kappa_2$& $-g_s^2\fpi^{2}/(2m_s^2)$           
&$-2.064$  &$-1.471$  & $-1.192$ \\
$\wt\zeta_2$& $g_v^2\fpi^{2}/(2m_v^2)$             
& $1.563$  & $0.948$ & $0.667$  \\
$\wt\xi_2$ & $g_\rho^2\fpi^{2}/(2m_\rho^2)$        
& $0.263$  & $-0.048$  & $0.364$  \\
$\wt\eta_1$&                    
&$-2.519$  & $1.172$  & $1.987$ \\
$\wt\eta_2$&    
&          &$-1.885$  &$-2.445$ \\
$\wt\kappa_3$&                 
& $3.209$  & $0.114$  &$-0.501$ \\
$\wt\kappa_4$&  
&          &$-0.314$  &$-1.607$ \\
$\wt\zeta_4$ &        
&          &$-0.314$  &$0.792$  \\
$\wt\eta_\rho$&               
& $1.670$  & $4.597$  &$1.148$  \\
$\wt f_v$&                                
& $0.023$  & $0.142$  & $0.251$ \\
$\wt f_\rho$         &                              
& $2.888$  &  $3.552$ & $3.328$ \\ 
$\wt\kappa_d$& $-g_s^2\fpi^{2}\Lambda^{2}/(2m_s^4)$           
&$-3.066$  & $5.019$ & $4.675$ \\
$\wt\zeta_d$& $g_v^2\fpi^{2}\Lambda^{2}/(2m_v^4)$             
& $1.474$  & $-6.310$  & $-5.812$\\
${\wt\xi}_d$& $g_\rho^2\fpi^{2}\Lambda^{2}/(2m_\rho^4)$       
& $-1.456$ &$-4.954$  & $-3.632$\\
$\wt\alpha_1$         &                              
& $15.63$  & $-37.83$ &$-37.43$ \\ 
$\wt\alpha_2$         &                              
& $-12.62$ &  $38.14$ &$35.84$\\ \hline
$\wt\kappa_2/\wt\kappa_d$ & $m_S^2/\Lambda^2$ 
& $0.673$  & $-0.293$  & $-0.255$ \\
$\wt\zeta_2/\wt\zeta_d$ & $m_v^2/\Lambda^2$ 
& $1.060$  & $-0.150$  & $-0.115$ \\
$\wt\xi_2/\wt\xi_d$ & $m_\rho^2/\Lambda^2$ 
& $-0.180$  & $-0.009$  & $-0.100$ \\
$-\wt\kappa_2^2/(8\pi^2\wt\kappa_d)$ 
& $g_s^2/(16\pi^2)$                               
& $1.182$   & $-0.367$  &$-0.258$ \\
$\wt\zeta_2/(8\pi^2\wt\zeta_d)$ & $g_v^2/(16\pi^2)$    
& $1.408$   & $-0.121$  & $-0.065$ \\
$\wt\xi_2^2/(8\pi^2{\wt\xi}_d)$ 
& $g_\rho^2/(16\pi^2)$                            
& $-0.040$  & $-0.0004$ & $-0.031$ \\ \hline
$\chi^2$ &   					  
& $33$  &  $25$  & $23$\\
\end{tabular}
\label{tab:coeffsVA}
\end{table}

\def\rhop{\rho_{+}}
\def\rhom{\rho_{-}}

\begin{table}[tbp]
\caption{Optimal coefficients for selected parameter sets.}
\begin{tabular}{c|c|rrrr|rrrr}
   parameter & order\ \ & FZ2\ & VZ2\ & FA2\ & VA2\ 
                        & FZ4\ & VZ4\ & FA4\ & VA4\ \\ 
     \hline\hline
  $\opt{1}$ & $\wt\rhop^{2}$ & $-0.49$ & $-0.49$ & $-0.50$ & $-0.50$ &
                            $-0.50$ & $-0.52$ & $-0.51$ & $-0.52$ \\ 
  $\opt{3}$ & $\wt\rhop^{3}$ & $0.76$ & $0.60$ & $0.69$ & $0.69$ &
                            $1.27$ & $1.33$ & $1.29$ & $1.49$ \\ 
  $\opt{2}$ & $\wt\rhop\wt\rhom$ & $-2.65$ & $-3.46$ & $-3.63$ & $-3.63$ &
                             $-1.58$ & $-2.45$ & $-2.08$ & $-1.86$ \\ 
  $\opt{5}$ & $\wt\rhop^{4}$ & $0.00$ & $0.00$ & $0.00$ & $0.00$ &
                            $-1.61$ & $-2.53$ & $-2.06$ & $-3.26$ \\ 
  $\opt{4}$ & $\wt\rhop^{2}\wt\rhom$ & $2.12$ & $3.14$ & $4.04$ & $4.05$ &
                                 $0.39$ & $1.26$ & $0.40$ & $-1.16$ \\ 
  $\opt{6}$ & $\wt\rhop^{3}\wt\rhom$ & $0.00$ & $0.00$ & $0.00$ & $0.00$ &
                                 $-2.66$ & $-3.58$ & $-1.74$ & $-2.40$ \\ 
  $\opt{7}$ & $\wt\rhop^{2}\wt\rhom^{2}$ & $0.00$ & $0.00$ & $0.00$ & $0.00$ &
                                     $0.00$ & $0.00$ & $0.00$ & $0.00$ \\
                        \hline 
  $\optder{1}$ & $\wt\rhop^{2}$ & $-1.36$ & $-1.35$ & $-1.60$ & $-1.59$ &
                               $-1.34$ & $-1.31$ & $-1.38$ & $-1.14$ \\ 
  $\optder{2}$ & $\wt\rhop\wt\rhom$ & $-3.46$ & $-2.13$ & $-4.63$ & $-4.54$ &
                                $-2.38$ & $-0.11$ & $-2.90$ & $10.49$ \\
                        \hline 
  $\optalpha{1}$ & $\wt\rhop^{3}$ & $0.00$ & $0.00$ & $3.30$ & $3.01$ &
                                 $0.00$ & $0.00$ & $0.74$ & $-1.59$ \\ 
  $\optalpha{2}$ & $\wt\rhop^{2}\wt\rhom$ & $0.00$ & $0.00$ & $29.54$ & $28.25$
          & $0.00$ & $0.00$ & $17.51$ & $-73.30$ 

\end{tabular}
\label{tab:optvalues}
\end{table}

The optimizations were carried out with the Minuit software
package \cite{Minuit}.
The resulting data sets for fits at various orders of truncation 
are presented in Tables~\ref{tab:coeffsFZ}, 
\ref{tab:coeffsVZ}, \ref{tab:coeffsFA} and
\ref{tab:coeffsVA}.  
Results from parameter sets FZ$0$ and VZ$0$
show that truncation at second order in density
leads to a poor fit with experimental data
and yields a similar $\chi^2$ to the analogous parameter
set W1 in the meson model from Ref.~\cite{Tang}. 
A significant difference from the meson model arises at the next level
of truncation, however:
the inclusion of only a single optimal parameter beyond second order in 
density yields a significantly better fit than the full third-order
results of the meson model\cite{Tang}.
Incorporating all third-order terms, sets FZ$2$ and VZ$2$
are comparable to or better
than the best fourth-order  results of the meson model 
given in Ref.\cite{Tang}.
Because of the improvements in the optimization procedure, however, 
a quantitative comparison of
$\chi^2$ values from this work and from the meson-model
calculations of Ref.~\cite{Tang} might be misleading.
In a future investigation the meson model will be reanalyzed
to determine if the differences noted here
are due to the improvements in optimization or arise from
a greater flexibility of the point-coupling model in fitting
nuclear properties.

Beyond the lowest level of truncation, 
fits to the charge radii, d.m.s.\ radii, binding energies, and spin-orbit
splittings for each data set
are all quite good; they are essentially all at the relative accuracy
prescribed by the corresponding weights (see Ref.~\cite{Tang} for details).
The fits here with $i \ge 2$ are as good as or better than the best
fits in Ref.~\cite{Tang}.  
Also, the low-momentum behavior of the 
nuclear charge form factors is well reproduced and is comparable
to those of the meson model\cite{Tang}.
Since our focus here is on testing naturalness 
and the NDA expansion in the point-coupling
model, we leave a detailed comparison between meson and point-coupling
models to a future work.

\begin{figure}
\setlength{\epsfxsize}{5.0in}
\centerline{\epsffile{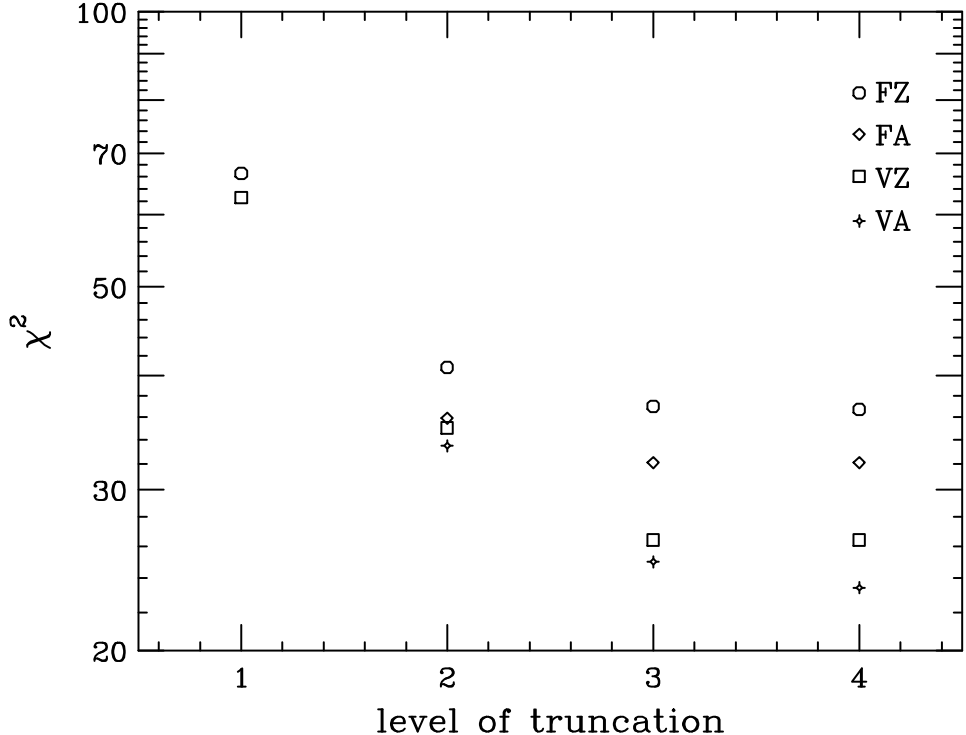}}
\caption{Best-fit $\chi^2$ values at different levels of truncation
     (see text).  Results for the lowest level of truncation are
     not shown.}
\label{fig:chi2}
\end{figure}

A plot of the value of $\chi^2$ for parameter sets
in each category at different levels of truncation is shown
in Fig.~\ref{fig:chi2}. 
The difference in $\chi^2$ between the third and fourth level of truncation
is small and is a signal that including higher-order terms 
will not improve the fit to data significantly.
This has been verified by optimizing fits for parameter
sets containing the leading fifth-order optimal parameter.
Also, at each level of truncation, parameter set VZ yields a better
fit than FA, and FZ and VA yield the worst and best fits, respectively.
That is, removing constraints on the parameters systematically
improves the fits.
Note however that {\it all\/} of the fits for $i>0$ are good!
We should be cautious, therefore, 
about drawing strong conclusions based on such small differences in the
$\chi^2$.

\subsection{Naturalness}

A review of the parameter sets FZ$i$ and VZ$i$ shows 
that most of the coefficients
are natural.  
For sets in which $\wt\alpha_1$ and $\wt\alpha_2$ are non-zero
(FA$i$ and VA$i$),
however, these parameters tend to be large and unnatural.
This tendency can be understood by considering the optimal
parameters defined earlier.
They are given in Table~\ref{tab:optvalues} for sets with $i=2$ and
$i=4$.
The second column indicates the leading power of $\wt\rhop$ and $\wt\rhom$
for the corresponding term.  

The coefficients are ordered according
to the empirical rule that $\wt\rhom$ should
be counted roughly as $\wt\rhop^{8/3}$ (near equilibrium density).
We expect that lower order parameters will be better determined
by the fit to the data, and this is indeed the case.
(In general, the optimal parameters are much better determined than the
original covariant parameters.)
Furthermore, when a set is truncated, 
the highest-order nonzero
coefficients are forced to compensate for the omitted terms.
In sets with $i=3$, the values of $\opt{1}$, $\opt{2}$,
$\opt{3}$ and $\opt{4}$ are consistent with the same range in values
as the sets with $i=4$.  Truncation of the term
corresponding to $\opt{6}$ can be compensated
by $\opt{5}$ (which is underdetermined at $i=4$ anyway).

The truncation at $i=2$ is different, however.
These sets give quantitatively good fits to the data,
but although the fits yield consistent values of
$\opt{2}$ and $\opt{3}$ for these sets, their
values are significantly different from those in sets with $i=3$ and $i=4$.
In fact, the large size of $\opt{2}$ compared to $\opt{3}$ for 
$i=2$ indicates a breakdown of the systematics: the 
contribution to the binding energy from 
the term corresponding to $\opt{2}$ is larger than for $\opt{3}$,
while the opposite behavior is expected.
Thus, although the $i=2$ level of truncation yields
good fits, it is clear that a higher level of truncation
is necessary for consistency and to produce the systematics
implied by naturalness.

From the table, it is clear that
the optimal parameter $\wt A_2$
is unnatural, while
the optimal parameter $\wt A_1 = \wt\alpha_1+\wt\alpha_2$
is of order unity in magnitude.
A similar situation exists for
$\wt\Delta_2 = 2(\wt\kappa_d-\wt\zeta_d)$ and 
$\wt\Delta_1 = \wt\kappa_d+\wt\zeta_d$: 
whereas the parameter $\wt\Delta_1$ seems to maintain a consistent
value on the order of $-1.3$ throughout the sets, the value
of $\wt\Delta_2$ varies wildly.
The combination $\wt\Delta_1 + \wt A_1 \langle\scalarN\rangle$,
where $\langle\scalarN\rangle$ is an average value  (roughly 1/10)
is even more tightly determined.
Since $\wt A_2$ is poorly determined by the observables, it is not
surprising that a fit could lead to individual values for $\wt\alpha_1$ 
and $\wt\alpha_2$ that are unnatural.

The point-coupling model is evaluated here for naturalness in the 
absence of any combinatorial counting factors (as in Ref.~\cite{Madland}).  
This
is in contrast to the meson model analysis, where a term with 
$n$ powers of the scalar or vector field 
incorporated factors of $1/n!$ in the treatment of naturalness \cite{Tang}.
Yet each analysis yielded natural coefficients in essentially
all cases.
We have not yet understood this difference; 
a reanalysis of the meson models using
the improvements to the minimization procedure discussed here is needed
before drawing firm conclusions.

\subsection{Nuclear Matter}

Saturation properties of nuclear matter for each of the parameter
sets are given in Table~\ref{tab:sat}.  The saturation point and 
binding energy per nucleon have values characteristic
of relativistic mean-field meson models 
that successfully reproduce bulk properties of nuclei \cite{FURNSTAHL96}.
The consistency of the numbers for these sets and the ones in
Ref.~\cite{Tang} is striking.
Different weights might shift the numbers slightly, but
within small errors the binding energy per nucleon is 16\,MeV and
the Fermi momentum is 1.3\,fm$^{-1}$.
The compressibilities are of order 300\,MeV for the more complete
models, 
which is somewhat high compared to the best fits from Ref.~\cite{Tang}.

\begin{table}[tbp]
\caption{Nuclear matter equilibrium properties}
\begin{tabular}{c|c|c|c|c|c|c}
Set  & $E/B-M$   &$\kfermi$&  $K$  & $a_4$  & $M^*/M$ & Vector Potential/$M$\\
\hline
FZ$0$& $-16.1$  & $1.27$ & $570$ & $39.4$ & $0.533$ & $0.387$ \\
FZ$1$& $-16.0$  & $1.30$ & $390$ & $32.6$ & $0.697$ & $0.234$ \\
FZ$2$& $-15.9$  & $1.30$ & $320$ & $31.8$ & $0.676$ & $0.253$ \\
FZ$3$& $-15.9$  & $1.31$ & $300$ & $29.9$ & $0.742$ & $0.191$ \\
FZ$4$& $-15.9$  & $1.31$ & $300$ & $30.0$ & $0.743$ & $0.190$ \\ \hline
FA$2$& $-16.0$  & $1.30$ & $290$ & $33.5$ & $0.601$ & $0.322$ \\
FA$3$& $-16.0$  & $1.31$ & $280$ & $29.7$ & $0.677$ & $0.252$ \\
FA$4$& $-16.0$  & $1.31$ & $290$ & $29.7$ & $0.681$ & $0.248$   \\ \hline
VZ$0$& $-16.3$  & $1.29$ & $560$ & $37.9$ & $0.535$ & $0.383$ \\
VZ$1$& $-16.0$  & $1.30$ & $380$ & $32.8$ & $0.698$ & $0.234$ \\
VZ$2$& $-16.0$  & $1.30$ & $330$ & $34.8$ & $0.597$ & $0.327$ \\
VZ$3$& $-16.0$  & $1.30$ & $300$ & $34.6$ & $0.623$ & $0.303$ \\
VZ$4$& $-16.0$  & $1.31$ & $310$ & $32.8$ & $0.645$ & $0.282$ \\ \hline
VA$2$& $-16.0$  & $1.30$ & $290$ & $35.0$ & $0.602$ & $0.322$ \\
VA$3$& $-16.1$  & $1.30$ & $300$ & $39.3$ & $0.629$ & $0.297$ \\
VA$4$& $-16.1$  & $1.30$ & $320$ & $34.5$ & $0.659$ & $0.269$ \\      
\end{tabular}
\label{tab:sat}
\end{table}

The
scalar and vector potentials are in many cases much smaller than typical
values found in meson models \cite{FURNSTAHL96}.
Previous experience with meson models implied that large
potentials were needed to reproduce experimental spin-orbit splittings.
The point-coupling
model of Ref.~\cite{Madland}, with $\Mstar/M$ of 0.58, was 
consistent with this result.  

The new feature here that allows for this deviation from
conventional wisdom is the isoscalar vector-tensor
coupling proportional to $\wt f_v$, the analog of which has not usually been
included in relativistic mean-field models%
\footnote{We note that the parameter $f_v$ in the meson model 
plays a role in determining electromagnetic properties through
VMD effects, whereas $\wt f_v$ in the point-coupling
model does not have any direct association with such effects.}
(often based on arguments
from vector dominance and the smallness of the isoscalar anomalous
moment of the nucleon).
It provides, however, an independent contribution to the spin-orbit
force.
In Ref.~\cite{Tang}, where the analog of this term 
(proportional to $f_v$) was included,
the optimization did not lead to 
particularly large
values of the tensor coupling $f_v$, 
although for set G2 it was large enough to drive the value of 
$\Mstar/M$ to 0.66.
This value is outside the range $0.58 <\Mstar/M < 0.64$ for conventional
models with good fits to nuclei \cite{FURNSTAHL96}.
In the point-coupling model there is apparently a significant advantage
to having larger $\wt f_v$ 
(at least with our choice of observables and weights)
and we find $\Mstar/M$ as large as 
0.74!

Since we are primarily concerned here with testing naturalness
and the NDA expansion and not with generating parameter sets for
practical use,
we have simply checked that setting $f_v = 0$ and re-optimizing the
parameters yields a
smaller value for $\Mstar$ (consistent with conventional wisdom of
nuclear models) and does not change any of our conclusions.
This is indeed the case.
A more complete discussion of the spin-orbit force and tensor couplings
in chiral effective field theories will be
given elsewhere \cite{Spinorbit}.

\begin{figure}[t]
\setlength{\epsfxsize}{4.0in}
\centerline{\epsffile{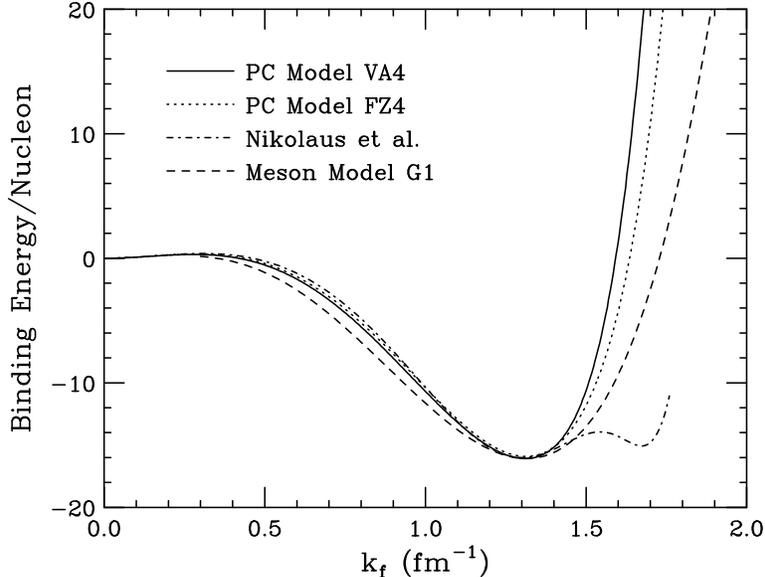}}
\caption{Nuclear saturation curves for point-coupling models
  VA4 and FZ4 from this work, the point-coupling model from
  Nikolaus et al.~\protect\cite{Madland}, and the meson model
  G1 from Ref.~\protect\cite{Tang}.}
\label{fig:saturation}
\end{figure}

In Fig.~\ref{fig:saturation},
the saturation curves for sets FZ$4$ and VA$4$ are plotted
against the results of set G$1$ from the meson model of Ref.~\cite{Tang}
and the earlier point-coupling results of Ref.~\cite{Madland}.
The larger compressibilities of the point-coupling models is
evident and leads to deviations from the meson-model result.
With increasing density, the energies for point-coupling models VA4 and FZ4
(and all others fit here) increase smoothly, as found for meson models
(such as G1).
In contrast, the point-coupling model 
found by Nikolaus et al.~\cite{Madland} shows a peculiar second
minimum.
This structure comes from
the contributions from the
highest-order terms (fourth power of density) 
in the model of Ref.~\cite{Madland}, which
are much larger than those of the next-to-leading order.
Their contributions to the value of $M^*/M$ and to the binding energy
become dominant
for values of the Fermi momentum not far from saturation,
indicating that the expansion has broken down;
for this restricted class of models 
the extrapolation in density is not valid.
The corresponding parameters are individually unnatural and there
are no mixed terms, so the $M^*$ equation is strongly affected.

\begin{figure}[p]
\setlength{\epsfxsize}{4.0in}
\centerline{\epsffile{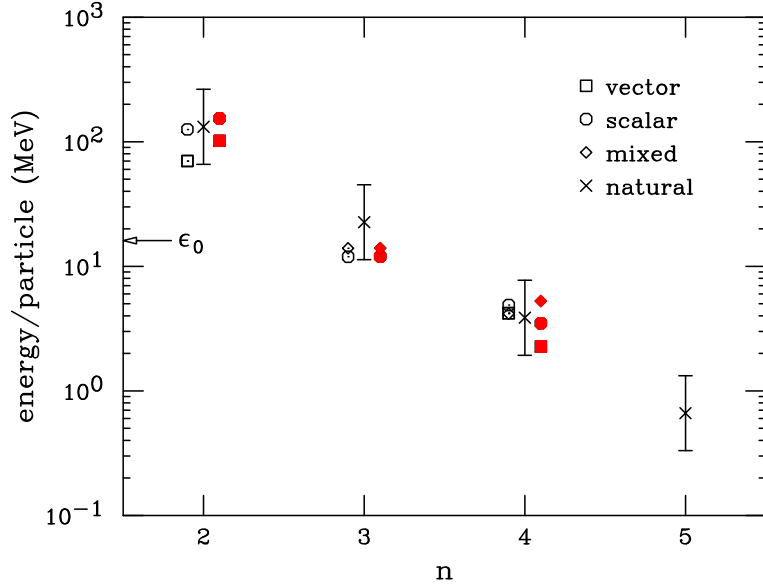}}
\caption{Contributions to the binding energy of equilibrium
nuclear matter from terms of
the form $\rho_s^{i}\rho_v^{j}$ with $i+j=n$. The symbols
indicate terms with $i=0$ (squares), $j=0$ (circles), and $i\neq0$, 
$j\neq 0$ (diamonds).  The unfilled symbols indicate set FZ4 and the
filled symbols indicate set FA4.  The crosses are estimates based
on Eq.~(\ref{eq:estimate}).}
\label{fig:be1}
\end{figure}

\begin{figure}[p]
\setlength{\epsfxsize}{4.0in}
\centerline{\epsffile{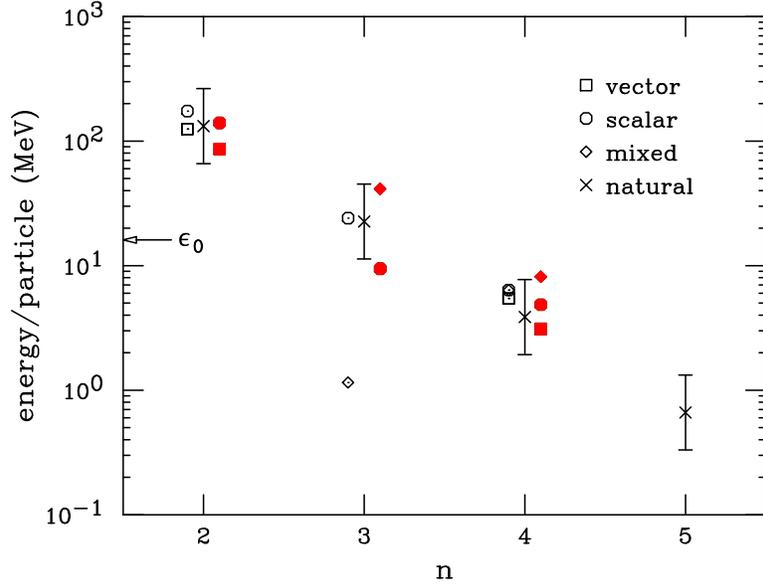}}
\caption{Contributions to the binding energy of equilibrium
nuclear matter from terms of
the form $\rho_s^{i}\rho_v^{j}$ with $i+j=n$. The symbols
indicate terms with $i=0$ (squares), $j=0$ (circles), and $i\neq0$, 
$j\neq 0$ (diamonds).  The unfilled symbols indicate set VZ4 and the
filled symbols indicate set VA4.  The crosses are estimates based
on Eq.~(\ref{eq:estimate}).}
\label{fig:be2}
\end{figure}

The implications of naturalness are clear in Figs.~\ref{fig:be1} and
\ref{fig:be2}, where
contributions to the binding energy per nucleon at equilibrium
from terms of the
form $\rho^i_s \rho^j_v$ with $i+j=n$ and $\rho_v \equiv j_v^0$ 
are plotted for various parameter sets. 
(The reader is cautioned not to confuse $n$ with the label of
the parameter sets
associated with the optimal hierarchy).
The scale of the natural size of contributions expected from
an $n$-th order term is obtained through NDA; since the difference
in the scalar and vector densities is small, (the magnitude of)
the contribution from
such a term can be estimated by
\begin{eqnarray}
\alpha_{\rm nat}\Lambda \left({\rho_v\over f_\pi^2\Lambda}\right)^{n-1}\ ,
\label{eq:estimate}
\end{eqnarray}
where $\alpha_{\rm nat}$ is a  (positive) number of order unity
and $\rho_v$ is the nuclear matter equilibrium density.
Estimates for $\Lambda = 770\,$MeV 
are marked in the figures with crosses and the error bars reflect a
range of $\alpha_{\rm nat}$ from $1/2$ to 2.
We observe that the estimates are quite consistent with the
energies found from the fits.

Since the contributions decrease steadily with increasing powers
of the densities, the expansion and truncation scheme proposed
under NDA is justified.
Note that this decrease will become more gradual as the density
is increased above equilibrium density.
Based on the estimates, a truncation at $n=4$ for ordinary
nuclear observables yields an error of order 1~MeV or less.
The contributions at fourth-order in the density
are on the order of the nuclear-matter binding energy, however, and
consistency implies we need to include these terms. In practice, however,
the $n=3$ truncation (parameter set P2)
is able to absorb the $n=4$ contributions.
This result is consistent with a recent analysis of the nonrelativistic
Skyrme energy functional \cite{Hackworth}.
On the other hand,
this was not the case in Ref.~\cite{Tang}, where the $n=3$ fit, while
quite reasonable, was also noticeably inferior to $n=4$.

While the expansion and truncation scheme is supported by
our results, a comparison of the various parameter sets shows 
significant variation for individual parameters.  In many cases even
the sign is indeterminate.  Thus the parameters in this approach,
as in the relativistic meson model, are underdetermined.  On the other
hand, Table~\ref{tab:optvalues} shows that the optimal parameters are
much better constrained by the data.
Therefore reformulating the effective theory in terms of an optimal basis
may be more productive.

\subsection{Equivalent meson masses and Yukawa couplings}

No restrictions were imposed during optimization
on the signs of the coefficients $\kappa_2$, 
$\zeta_2$, and $\xi_2$ nor on 
$\kappa_d$, $\zeta_d$, and  $\xi_d$ when they were
allowed to vary (the ``V'' sets).
On the other hand, as shown in Tables~\ref{tab:coeffsFZ}--\ref{tab:coeffsVA},
a direct transformation from a relativistic mean-field meson model
would predict definite signs for these coefficients; the coefficients
determined through this transformation would
depend only on the squares of the Yukawa couplings, $g_i^2$, as well
as the squares of the meson masses, $m_i^2$.
Sets FA$n$ and FZ$n$ all have positive values for the corresponding
values of $g_i^2$ as well as for the scalar mass squared.
(Recall that the corresponding values of 
$m_v$ and $m_\rho$ were held fixed for these sets).
Parameters sets VZ$n$ and VA$n$, however, 
include sets with $m_i^2 < 0$ and $g_i^2 < 0$.
The ratio $g_i^2/m_i^2$, which is the integrated
strength in each channel, is reasonably well determined and is in
every case positive.

The difficulty with predicting individual masses and couplings is revealed
by Table~\ref{tab:optvalues}.  While $\optder{1}$ is well determined
and consistent across the parameter sets, $\optder{2}$ is poorly determined.
Thus, the individual values for $\wt\kappa_d$ and $\wt\zeta_d$ are also
poorly determined, and we cannot reliably extract masses and couplings
from the fits.
We conclude that the nuclear observables we have used do not
provide sufficient constraints
to definitively test whether the point couplings are dominated by
an underlying meson phenomenology.
The combinations that {\it are\/} well determined correspond to the strength
and range of the effective central potential.  
Specifically, if the static nonrelativistic potential in momentum space
from the exchange of scalar and vector mesons is written as:
\begin{equation}
   V({\bf q}) = 4\pi \left(-{g_s^2 \over m_s^2 + {\bf q}^2}
            + {g_v^2 \over m_v^2 + {\bf q}^2} \right) \ ,
\end{equation}
and expanded in powers of ${\bf q^2}$,
\begin{equation}
  V({\bf q}) = 4\pi \left(-{g_s^2 \over m_s^2}+ {g_v^2 \over m_v^2}\right)
        - 4\pi {\bf q^2} \left(-{g_s^2 \over m_s^4}+ {g_v^2 \over m_v^4}\right)
       + \cdots   \ ,
\end{equation}
the two combinations of couplings and masses in parentheses are 
reasonably well determined.

\section{Summary}

The successful application of EFT concepts to mean-field mesonic models of 
nuclei \cite{Tang}
motivates their application to point-coupling models.  Such models
describe nuclear interactions through contact terms (in a derivative
expansion) in place of 
meson exchange.  In comparison to the large body of work 
on meson models, 
the analysis of relativistic point-coupling models has been quite limited.
Here, we extend the model of Nikolaus, Hoch, and Madland 
\cite{Madland} and its analysis by Friar et al.~\cite{Friar} 
to encompass a more complete analysis
based on EFT concepts.

The lagrangian is consistent with the symmetries of QCD
and is organized according to the same principle
of naive dimensional analysis (NDA) applied to the relativistic
meson model of Ref.~\cite{Tang}.
The organization is an expansion 
in powers and derivatives
of the densities in ratio
to  scales dictated by NDA results: a given term in the 
lagrangian takes the form
\begin{eqnarray}
c\left[\left({\partial\over \Lambda}\right)^p
	\left({\overline N \Gamma N\over f_\pi^2\Lambda}\right)^l\right]
\end{eqnarray}
where $c$ is a dimensionless coefficient and $p$ and $l$ are integers.
(Electromagnetic interactions
will also contain a power of the photon field and the electric charge).
In principle, all terms consistent with the symmetries should be 
included to a given order in the expansion, but in the present
work we have omitted a
variety of terms that we expect to be poorly determined based on
meson exchange phenomenology and experience with relativistic meson
mean-field models.

The NDA organization provides a valid expansion scheme provided the
coefficients are ``natural'' (on the order of unity); one can construct
an energy functional in powers of densities and their derivatives and 
truncate at some finite order with a controlled error.  
Our fits to bulk nuclear 
properties show this to be the case.
Beyond a second-order
truncation, the models resulting from
these fits reproduce the experimental data quite well.
We conclude that NDA and the naturalness assumption are compatible with
and implied by the observed properties of finite nuclei, even though
many-body effects are absorbed into the coefficients.

Point-coupling models therefore provide an alternative phenomenology
to mean-field meson models.  In principle, a direct transformation
exists between any point-coupling and meson mean-field model.
The equations of motion for the mean meson fields can be iterated
to solve the fields as an expansion in scalar and vector densities.
The delicate cancellations at low order in the expansion result in 
too large a truncation error for such transformations to be of use,
however, and 
{\it any point-coupling model derived from a mesonic model in this way
must be refit to the data\/}.  
We note that naturalness of the point-coupling model was found
in the absence of counting factors that were required in the meson
model analysis. 

Although fourth-order terms would be required for a consistent
truncation, a truncation
at third order in the point-coupling model can yield a good
fit to data, in contrast to the meson models where fourth-order
terms were necessary in obtaining a good fit.
Due to improvements in the optimization procedure over that
used in Ref.~\cite{Tang} through a reorganization
of the lagrangian in terms of ``optimal parameters'', 
a true comparison cannot be made until the fits obtained in that
reference are reanalyzed.

The analysis in terms of optimal parameters suggests that they may
provide a more efficient basis for an expansion.
While the individual coefficients in the covariant point-coupling model are
in general poorly determined by the nuclear data, the lower-order
optimal parameters are quite well constrained and only the highest-order
parameters are badly underdetermined.  
The use of optimal parameters was suggested by methods of the
heavy baryon formulation of chiral perturbation theory, adapted to
finite density.
Further development of this approach and its relation to
nonrelativistic energy functionals for nuclei is in progress.


\acknowledgments

We thank S.~Brand, J.~Hackworth, and B.~Serot for useful discussions.
This work was supported in part by the 
National Science Foundation
under Grants No.\ PHY--9511923 and PHY--9258270.

\end{document}